\documentclass[aps,prb,superscriptaddress,amsmath,amssymb,reprint]{revtex4-2}

\usepackage{tikz}
\usepackage{circuitikz}
\usepackage{graphicx}
\usepackage{dcolumn}
\usepackage{bm}
\usepackage{svg}
\usepackage{braket}
\usepackage{bbold}
\usepackage{enumerate}
\usepackage{wrapfig}
\usepackage{amsmath} 
\usepackage{amssymb} 
\usepackage{seqsplit}
\usepackage{xcolor}
\usepackage{braket}
\usepackage{tabularx}
\usepackage{makecell}
\usepackage[export]{adjustbox}
\usepackage[shortlabels]{enumitem}
\usepackage{setspace}
\usepackage{tensor}
\usepackage{hyperref}
\usepackage[capitalise]{cleveref}
\usepackage{float}
\usepackage{comment}
\usepackage{multirow}
\usepackage{longtable}

\newcommand{\bes} {\begin{subequations}}
\newcommand{\ees} {\end{subequations}}

\def\theYear{\the\year}
\graphicspath{ {./figures/} }
\hypersetup{pdfpagemode=FullScreen, colorlinks=true, allcolors=blue}

\begin{document}

\title{A Review of Design Concerns in Superconducting Quantum Circuits}
\author{Eli M. Levenson-Falk}
\email{elevenso@usc.edu}
\affiliation{Center for Quantum Information Science and Technology, University of Southern California, Los Angeles, California 90089}
\affiliation{Department of Physics \& Astronomy, University of Southern California, Los Angeles, California 90089}
\affiliation{Ming Hsieh Department of Electrical \& Computer Engineering, University of Southern California, Los Angeles, California 90089}

\author{Sadman Ahmed Shanto}
\affiliation{Center for Quantum Information Science and Technology, University of Southern California, Los Angeles, California 90089}
\affiliation{Department of Physics \& Astronomy, University of Southern California, Los Angeles, California 90089}

\begin{abstract}
In this short review we describe the process of designing a superconducting circuit device for quantum information applications. We discuss the factors that must be considered to implement a desired effective Hamiltonian on a device. We describe the translation between a device's physical layout, the circuit graph, and the effective Hamiltonian. We go over the process of electromagnetic simulation of a device layout to predict its behavior. We also discuss concerns such as connectivity, crosstalk suppression, and radiation shielding, and how they affect both on-chip design and enclosure structures. This paper provides an overview of the challenges in superconducting quantum circuit design and acts as a starter document for researchers working on any of these challenges.
\end{abstract}

\maketitle
\tableofcontents

\section{Introduction}\label{sec:intro}

Superconducting circuits are a leading quantum computing technology, combining strong couplings and long-lived coherence with flexible engineering and scalability \cite{kjaergaardSuperconductingQubitsCurrent2020}. By carefully arranging of capacitors, inductors, transmission lines, and Josephson junctions, experimenters create circuits with coherent quantized energy levels, often called \emph{artificial atoms}. While every atom of a particular isotope is completely identical, artificial atoms are extremely customizable. Their behavior is determined by the \emph{circuit graph}, which shows the arrangement of each circuit element and the connections between them; the \emph{circuit element parameters}, i.e., the exact values of the inductances, capacitances, impedances, antenna couplings, etc.; the \emph{layout}, i.e., the physical geometry of superconducting metal, insulating substrate, dielectric layers, etc.; the \emph{embedding structure}, i.e., the physical enclosure that houses a circuit and routes signals to it; and the \emph{materials} used in fabricating the circuit as well as the fabrication processes that affect surface and interface layers. Each of these aspects must be carefully engineered to create the desired behavior.

The \emph{design problem} in superconducting quantum devices can be stated succinctly: given some desired behavior, how can we create a physical device that will produce this behavior? In this short review we discuss the various concerns in designing superconducting quantum circuits.

A typical design process may be summarized in steps:
\begin{enumerate}
    \item Determine the desired behavior, i.e., the desired Hamiltonian.
    \item Find a circuit model that implements this Hamiltonian.
    \item Create a physical device layout that the designer guesses will implement the circuit model.
    \item Perform finite-element electromagnetic simulations of the layout and construct a circuit model and/or Hamiltonian based on the simulation results.
    \item Compare the model/Hamiltonian from the simulation to the desired Hamiltonian. If they do not match to within the required tolerance, alter the layout, simulate, compare, and iterate steps 3-5 until the predicted behavior matches the desired behavior.
    \item Fabricate the device, measure its behavior, and compare to the simulation output (which matches target behavior). If experiment results deviate significantly from simulation, refine the simulation and/or fabrication pipelines and repeat steps 3–6.
\end{enumerate}

The review is structured according to this design loop workflow. It begins with a brief description of the physics of superconducting quantum circuits at an effective Hamiltonian level. We then discuss depictions of these circuits as being composed of lumped elements in a particular circuit graph, and how these depictions can be used to design the desired effective Hamiltonian. We follow this by discussing how a physical device layout is designed and how the layout can effect behavior. We then discuss simulations that can be used to predict device behavior given a layout. We describe how the device can change between layout and fabrication. And finally we close the loop by describing how measured device behavior can be used to improve the next round of designs. Throughout each discussion we describe how the steps interact with each other. The goal of this review is not to provide a comprehensive, detailed description of each aspect of design---in many cases these articles have already been written, and we refer the reader to them throughout the text. Instead this review is intended as an overview that can be used as an introduction to superconducting device design for new researchers in the field, and a centralized reference for locating more detailed resources on each aspect of the problem. \cref{tab:references} summarizes these references, organized by topic.


\section{Superconducting Qubits} \label{sec:SCQ}
Here we give a brief introduction to the behavior of superconducting circuits. For more detailed discussions, see references \cite{clarkeSuperconductingQuantumBits2008,devoretSuperconductingCircuitsQuantum2013a,wendinQuantumInformationProcessing2017a,krantzQuantumEngineersGuide2019,kjaergaardSuperconductingQubitsCurrent2020,kwonGatebasedSuperconductingQuantum2021,blaisCircuitQuantumElectrodynamics2021a,gaoPracticalGuideBuilding2021,rasmussenSuperconductingCircuitCompanionan2021}.

\begin{figure}[h]
\centering
    \includegraphics[width=\linewidth]{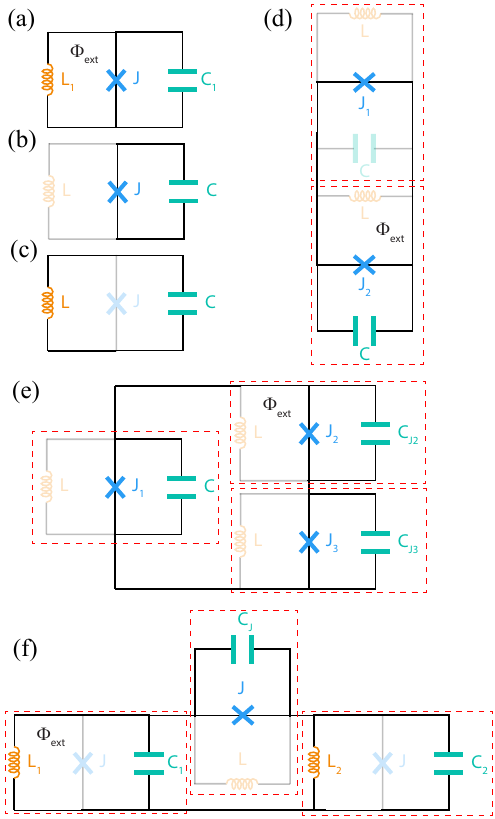} 
    \caption{\label{fig:LJC} (a) Diagram of an inductor (left, orange), Josephson junction (center, blue), and capacitor (teal, right), forming a parallel LJC circuit. An external flux $\Phi_\mathrm{ext}$ threads the loop formed by the L and J. Each element has an associated energy. Tiling LJC subcircuits in series and parallel allows one to build up an arbitrary circuit, including the fixed-frequency transmon (b), harmonic oscillator (c), tunable transmon (d), 3-junction flux qubit (e), and inductively-shunted transmon (f). In examples (b–f) we gray out any circuit element that is omitted, effectively setting its energy to 0, and draw dashed red lines around the LJC subcircuits that combine to make up the overall circuit.}
\end{figure}

\subsection{Quantizing a circuit}
A circuit can be modeled as a set of \textit{nodes} (each with its own voltage) connected by \textit{branches} (each carrying its own current). A branch can be formed by any lumped circuit element. Neglecting loss and quantum phase slips (see \cref{sec:graphtoH}), in a superconducting circuit such an element can be an inductor (with inductance $L$), a Josephson junction (with critical current $I_0$), or a capacitor (with capacitance $C$) (we neglect distributed elements for the moment). One can combine these into the \emph{parallel LJC} subcircuit. See \cref{fig:LJC} for a diagram. Note that the capacitance here may be entirely due to the Josephson junction geometry or may have some additional contribution from a physical shunting capacitor, and the inductance is a physical shunting inductance in parallel with the junction. This is the simplest possible circuit that captures all possible branches between two nodes and yields the following Hamiltonian
\begin{equation}
    \hat{H} = 4 E_c \left(\hat{n}-n_g\right)^2 - E_J \cos \hat{\phi} + \frac{E_L}{2}\left(\hat{\phi}-\phi_{ext}\right)^2.
\end{equation}
Here $\hat{n}$ is the charge number operator (where charge is quantized in units of Cooper pairs, i.e., $2e$), $n_g$ is a constant that accounts for the effect of voltage bias across the capacitor, $\hat{\phi}$ is the dimensionless node flux (that is, the phase difference across the junction/inductor with one side defined as 0---see \cref{sec:graphtoH} for a rigorous definition), and $\phi_{ext}$ is the dimensionless external flux threading the junction-inductor loop $\phi_{ext} = \frac{\Phi_{ext}}{\varphi_0}$, where $\varphi_0 \equiv \hbar/2e$ is the reduced magnetic flux quantum. Note that the node flux differs from the loop flux threading the circuit loop, as explained in \cref{sec:graphtoH}. The charge number operator and node flux obey the canonical conjugation relation $\left[\hat{\phi},\hat{n}\right]=i$ similar to position and momentum. The behavior of this subcircuit can be entirely captured by the component energies $E_c \equiv e^2 / 2C$, $E_J \equiv \varphi_0 I_0$, and $E_L \equiv \varphi_0^2/L$, along with the environmental bias variables $n_g$ and $\phi_{ext}$. A time-dependent $\phi_{ext}$ can lead to another term, but we will ignore this for simplicity \cite{youCircuitQuantizationPresence2019,riwarCircuitQuantizationTimedependent2022,bryonTimeDependentMagneticFlux2023}.

While the parallel LJC subcircuit need not be a qubit, one can produce certain qubit types from a single LJC circuit by varying the energies. This includes the ubiquitous transmon ($E_L = 0$, $E_J \gg E_c$)  and the many varieties of fluxonium ($E_c \sim E_J \gg E_L$), as well as harmonic oscillators ($E_J = 0$). Subcircuits such as this can be placed in series or parallel to generate an arbitrary lumped element circuit---omission of an element can be represented as setting its energy to 0. Examples of common simple combinations are shown in \cref{fig:LJC}(b), with omitted elements grayed out but still shown to emphasize this point (note that in all other figures we do not draw elements which are absent from the circuit, as is typical). Recent work has enumerated all possible combinations of such 2-node building blocks into circuits with up to 5 nodes \cite{weisslerEnumerationAllSuperconducting2024}, providing pre-solved Hamiltonians for such circuits. General circuit Hamiltonians can be built up as described in \cref{sec:graphtoH}. We have omitted loss (i.e., resistance) from this picture, which we will discuss later in \cref{sec:interfaces}.

\subsection{Quantum vs. classical descriptions}
This description assumes the Josephson junction itself is an ideal Josephson element with no resistive transport and only a single cosinusoidal Hamiltonian contribution $\hat{H}_J=-E_J\cos\hat{\phi}$. This is equivalent to the canonical classical model of a Josephson junction, the resistively capacitively shunted junction (RCSJ) model \cite{tinkhamIntroductionSuperconductivity2004}, with two modifications. The first is to take the limit $R\rightarrow\infty$. This limit is appropriate as superconducting quantum circuits are operated deep in the superconducting limit where normal-state transport is negligible. The second modification is more fundamental: the RCSJ model treats the junction as an AC voltage source and treats voltage and current (i.e., charge and flux) as classical coordinates. A full quantum treatment of the junction treats charge and flux as operators to correctly account for the fact that they do not commute, and uses the quantum Hamiltonian to compute state evolution.

Some numerical and theoretical work has demonstrated that the clssical RCSJ model can predict some behavior that had previously been described as purely quantum \cite{blackburnSurveyClassicalQuantum2016}, even including early entanglement demonstrations \cite{gronbech-jensenTomographyEntanglementCoupled2010}. It is also very useful for designing classical circuits such as amplifiers and as a starting point for quantum analysis to model the behavior of a composite circuit that functions as a single junction \cite{banszerusHybridJosephsonRhombus2025}. However, it cannot explain true quantum behavior, which at this point is well-established via experiments such as remote measurement-induced entanglement via one-way interactions \cite{rochObservationMeasurementinducedEntanglement2014}, Bell violations \cite{storzLoopholefreeBellInequality2023}, microwave-optical photon conversion \cite{mirhosseiniSuperconductingQubitOptical2020}, and the many demonstrations of quantum algorithms on superconducting processors. We therefore use a purely quantum treatment of the circuit, which has been fantastically successful at predicting observed behavior. 

On the other hand, the assumption of a cosinusoidal Hamiltonian may \textit{not} be valid, as recent evidence has shown that junctions may not behave as ideal tunnel junctions \cite{willschObservationJosephsonHarmonics2024}. Other junctions such as those using semiconducting nanowires \cite{haysCoherentManipulationAndreev2021,zhuoHoletypeSuperconductingGatemon2023}, 2D seminconductors \cite{sagiGateTunableTransmon2024,stricklandGatemoniumVoltageTunableFluxonium2025}, or other materials may also be used, which have intrinsically non-sinusoidal current-phase relations. One phenomenological approach is to approximate the junction behavior using a finite Fourier series. That is, one uses $N$ Hamiltonian terms of Josephson harmonics, replacing $-E_J\cos\hat{\phi}\rightarrow\sum_{m=1}^N -E_{J,m}\cos m\hat{\phi}$. 
This allows the Hamiltonian to be expressed using a finite number of well-behaved analytic terms.

\subsection{Effective Hamiltonian description}
\begin{figure}

\centering
    \includegraphics[width=\linewidth]{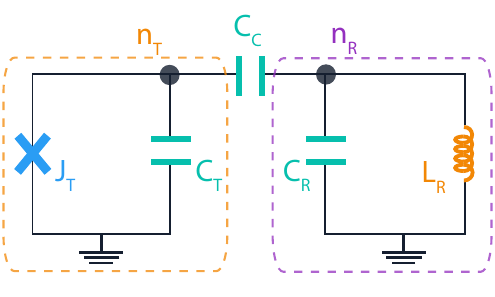} 
\caption{\label{fig:cQED} Circuit diagram of a transmon (left) capacitively coupled to a linear LC resonator (right). The node charges $n_T$ and $n_R$ are indicated by the nodes at which they are defined.}
\end{figure}

This circuit Hamiltonian description is accurate and, as we will see, is the level at which device designers must work. Unfortunately it is often inconvenient to work with when designing quantum architectures, algorithms, and gates, and for most quantum information science theory. Instead, the circuit Hamiltonian is typically reduced to an \emph{effective Hamiltonian} description, treating the circuit as a set of modes of oscillation which are best described as qubits, qudits, harmonic oscillators, and weakly anharmonic oscillators. These modes are coupled, typically with sufficiently weak couplings such that the original modes still form a useful basis. As an illustrative example, consider a single transmon coupled to a linear $LC$ resonator via a coupling capacitor as in \cref{fig:cQED}. Here and in all following diagrams we only draw circuit elements which are actually present, in contrast with the grayed out elements in \cref{fig:LJC}. 
Defining $E_{C,i} = e^2/2C_i$ and $E_L = \varphi^2/L_R$, the circuit-level Hamiltonian is
\begin{align}\label{eq:coupled-transmon}
    \hat{H} &= 4 E_{C,T} \hat{n}_T^2 - E_J \cos\hat{\phi}_T + 4 E_{C,R} \hat{n}_R^2 + \frac{1}{2}E_L\hat{\phi}_R^2 \\ \notag
    &+ 2 e^2 C_c\left(\frac{\hat{n}_T}{C_T}-\frac{\hat{n}_R}{C_R}\right)^2
\end{align}
where the $T$ and $R$ subscripts refer to transmon and resonator circuit elements and operators, respectively, and the charge number operators $\hat{n}_T$, $\hat{n}_R$ are the charge differences from the highlighted nodes to ground. In the language of LJC subcircuits this is the Hamiltonian of 3 subcircuits combined: a transmon (with zero inductive energy), a harmonic oscillator (with zero Josephson energy), and a coupling capacitor (with zero inductive or Josephson energy). 

This circuit can be described by an effective Hamiltonian:
\begin{align}\label{eq:cQED}
    \hat{H} &= \omega_R \hat{a}^\dag \hat{a} + \omega_T \hat{b}^\dag \hat{b} + \frac{\eta}{2}\left[(\hat{b}^\dag \hat{b})^2 - \hat{b}^\dag \hat{b}\right] \\ \notag
    &+ g(\hat{a} - \hat{a}^\dag)(\hat{b}-\hat{b}^\dag).
\end{align}
Here $\hat{a}$ is the lowering operator for the resonator mode, $\hat{b}$ is the lowering operator for the transmon mode, $\omega_i$ are the mode transition frequencies from ground to first excited state (when uncoupled), $\eta = \omega_{12,T}-\omega_{01,T}$ is the transmon anharmonicity, $g$ is a coupling strength between the modes, and we have set $\hbar = 1$. The mapping between circuit and effective Hamiltonian can be computed numerically, but is approximately
\begin{align} \label{eq:g}
&\omega_R = \sqrt{8 E_L E_{C,R}^\prime} \notag \\
&\omega_T \approx \sqrt{8 E_J E_{C,T}^\prime} - E_{C,T}^\prime \notag \\
&\eta \approx -E_{C,T}^\prime \notag \\
&g \approx \frac{C_c}{C_T}\sqrt{\frac{e^2\omega_r}{C_R}}\left(\frac{E_J}{8E_{C,T}}\right)^{1/4}
\end{align}
where $E_{C,i}^\prime = e^2/2(C_i+C_C)$. Note that these approximations are valid in the usual transmon limit $E_J>>E_{C,T}$ and the weak coupling limit $C_c<<C_T,C_R$ \cite{kochChargeinsensitiveQubitDesign2007}. 

A common simplifying assumption is to treat the transmon as a qubit and drop the non-excitation-conserving couplings $ab$ and $a^\dag b^\dag$ (i.e., take the rotating wave approximation, RWA), in which case one recovers the well-studied Jaynes-Cummings Hamiltonian \cite{shoreJaynesCummingsModel1993}:
\begin{equation*}
\hat{H} = \omega_R \hat{a}^\dag \hat{a} - \frac{\omega_T}{2} \hat{\sigma}_z - g(\hat{a}\hat{\sigma}_+ +\hat{a}^\dag \hat{\sigma}_-).
\end{equation*}
This depiction is conceptually simple, but is often not accurate enough in practice for, e.g., designing gate schemes \cite{scheuerPreciseQubitControl2014,willschGateerrorAnalysisSimulations2017,annSidebandTransitionsTwomode2021}, understanding measurement-induced state transitions \cite{sankMeasurementInducedStateTransitions2016a,dumasMeasurementInducedTransmonIonization2024}, or interpreting spectroscopic results \cite{zuecoQubitoscillatorDynamicsDispersive2009,Shanto2024squaddsvalidated} (see below). One message we hope the reader takes away from this review is: do not neglect the higher transmon levels or the non-RWA terms unless you have already confirmed they will not affect the result! This means that one must first calculate behavior including higher levels and non-RWA terms and compare to the behavior without. Then, if there is no significant difference, one could neglect the higher levels in any situation where the qubit approximation / RWA is expected to work even better. For higher levels, this means only using the qubit approximation in cases where the ratio of the relevant energy to the anharmonicity (e.g., the ratio of Rabi frequency to anharmonicity in a single-qubit gate, or the ratio of coupling strength to anharmonicity in a two-qubit coupling Hamiltonian) is even smaller than in the case where the approximation has been validated. For the RWA, this means only using it in cases where the ratio of difference between the relevant energies to their sum (e.g. the qubit and resonator if calculating energy shifts, or the drive frequency and qubit if calculating gate behavior) is even smaller than in the case where the approximation has been validated.

This circuit---a nonlinear mode that functions as a qubit, coupled to a linear harmonic oscillator---is the essential building block of circuit quantum electrodynamics (cQED), which is the architecture upon which almost all superconducting quantum circuits are based \cite{blaisCircuitQuantumElectrodynamics2021a}. Typically these circuits are implemented in the regime $|\Delta| \equiv |\omega_R-\omega_T| \gg |g|$. In this case one can diagonalize into resonator-like and transmon-like normal modes with self-Kerr (i.e., anharmonicity) terms $\eta_i$ and cross-Kerr (i.e., dispersive) terms $\chi$ in their energies:
\begin{align}\label{eq:diagKerr}
    \hat{H} &=  \omega_R^\prime \hat{a}^\dag \hat{a}  + \omega_T^\prime  \hat{b}^\dag \hat{b} \notag \\ 
    &+ \frac{\eta_R}{2} \left[(\hat{a}^\dag \hat{a})^2 - \hat{a}^\dag \hat{a}\right] + \frac{\eta_T}{2} \left[(\hat{b}^\dag \hat{b})^2 - \hat{b}^\dag \hat{b}\right] \notag \\ 
    &+ \chi \hat{a}^\dag \hat{a} \hat{b}^\dag \hat{b}.
\end{align}
This is often the description that maps best onto experimental spectroscopic measurement. Using such measurements, one can extract the parameters of the the original effective Hamiltonian \cref{eq:cQED} using analytical expressions for the energies (derived in 2nd-order perturbation theory):
\begin{align}
    \omega_R^\prime &\approx  \omega_R + g^2\left(\frac{1}{\Delta}-\frac{1}{\Sigma}\right) \\
    \omega_T^\prime &\approx  \omega_T - g^2\left(\frac{1}{\Delta} + \frac{1}{\Sigma}\right) \\
    \eta_R &\approx  0 \\
    \eta_T &\approx \eta  \\
    \chi &\approx  g^2 \left(\frac{\eta}{\Delta(\Delta-\eta)}+\frac{\eta}{\Sigma(\Sigma+\eta)}\right)
    \label{eq:kerr}
\end{align}
where $\Sigma \equiv \omega_R + \omega_T$. Note that the $\Sigma$ terms arise from counter-rotating (non-excitation-conserving) couplings such as $ab$ and $a^\dag b^\dag$, which are neglected under the RWA. However, they often can be significant, especially when computing the Lamb shift $\chi_L \equiv \omega_R^\prime - \omega_R$. A typical parameter regime is $\omega_R/2\pi \approx 7$ GHz, $\omega_T / 2\pi \approx 4$ GHz, and so omitting the $\Sigma$ term overestimates $\chi_L$ by approximately 25\%. Again, these expressions were derived perturbatively, and a fuller analytical or numerical treatment is required to capture features such as the small inherited resonator self-Kerr term $\eta_R$ \cite{zhangDriveinducedNonlinearitiesCavity2022}, the precise value of the transmon self-Kerr $\eta_T$ \cite{kochChargeinsensitiveQubitDesign2007}, and higher-order nonlinearities. These features are quite crucial when fine-tuning gate parameters or using the transmon to control a cavity qubit encoding \cite{krastanovUniversalControlOscillator2015,joshiQuantumInformationProcessing2021,maQuantumControlBosonic2021,krasnokSuperconductingMicrowaveCavities2024}. However, the goal of this section is merely to give an overview of the physics of these circuits, so we neglect these higher-order effects.

\section{Circuit Graph and Parameters} \label{sec:graph}

\begin{figure}
\begin{center}
    \makebox[\linewidth]{\hspace{10mm}\includegraphics[width=\linewidth]{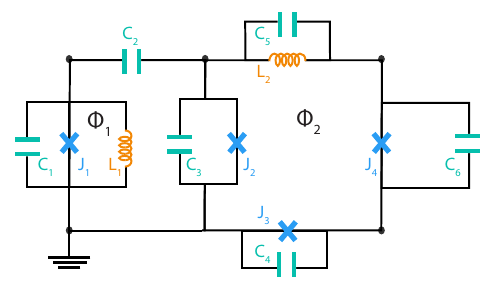}} 
    \caption{\label{fig:graph} An arbitrary circuit graph. Nodes (black dots) are connected by branches which may be inductive, capacitive, or Josephson elements. Each loop has an independent bias flux; each capacitive branch also has a bias voltage, which we omit for simplicity. From the set of capacitive branches one can define a spanning tree - effectively choosing a gauge for the flux variables - and thus assign node fluxes and charges accordingly.}
\end{center}
\end{figure}

It is possible to perform an electromagnetic simulation of a device layout and directly extract its energy spectrum and eigenstates, leading directly to an effective Hamiltonian description \cite{Minev2021} (see \cref{sec:em-qm} for an in-depth discussion). However, if one wishes to analytically derive the eigenstates and energies, develop a fundamentally new type of qubit, or sweep over a variety of device designs, such simulations may not be appropriate. Instead, researchers describe the device by its circuit graph, treating it as a set of lumped capacitors, inductors, and junctions between circuit nodes. A general example is shown in \cref{fig:graph}. This circuit graph description provides a useful middle step between the physical layout and the diagonalized Hamiltonian. This is an artificial creation, as only the device layout and the behavior it implements are actually physical. However, the circuit graph is such a useful tool that we will treat it as if it were physical.

\begin{figure*}  
    \centering
    \includegraphics[width=\textwidth]{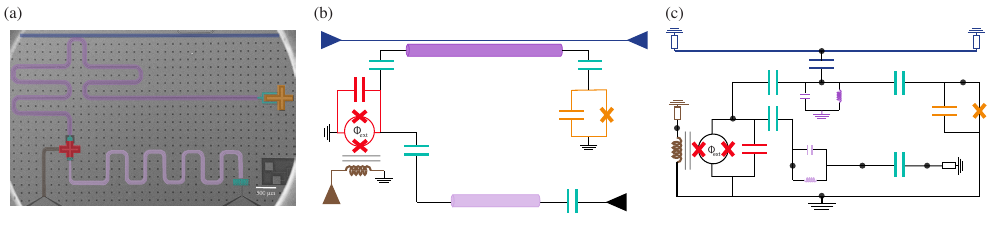}
    \caption{(a) Scanning electron microscope image of a superconducting device. False color highlights indicate two CPW resonators (purple and lilac), a tunable coupler (red), a fixed-frequency transmon (orange), and coupling capacitors (teal). The blue line at the top is a transmission line through, while the unhighlighted line to the left is a fast flux line inductively coupled to the coupler's SQUID. (b) Intermediate depiction of the device, showing circuit elements and their arrangement, with colors matching the coloring in (a). Distributed elements are shown as transmission lines. Full subcircuits, including the schematic lines connecting different elements to a node, are colored to show identification of subcircuits with their own resonant modes (c) Circuit graph of the device, with nodes indicated. Ports are changed to lumped-element resistors and distributed elements are replaced with effective lumped models. Elements are colored according to their subcircuits in (b), but connections within nodes are left black to indicate no pre-identification of modes \label{fig:layoutToGraph}}
    \label{fig:main}
\end{figure*}

\subsection{From Graph to Hamiltonian}\label{sec:graphtoH}

A circuit graph represents the device as if it were made of a finite number of nodes connected by branches, with each branch formed by a zero-size linear capacitor, linear inductor, or Josephson junction. More than one branch may connect a pair of nodes, and branches may couple to each other via mutual inductance. Again this is an approximation; in reality, every part of the device couples to every other part, and all device components have nonzero size. However, it is generally accurate enough to take the approximation that capacitances below a certain level, inductances above a certain level, and junction critical currents below a certain level can be neglected. Similarly, components whose dimensions are much smaller than the wavelength at the frequency considered can be treated as having zero size.

To see how to use the circuit graph, it is helpful to begin by treating the circuit purely classically. Each branch $b$ has a classical current $i_b(t)$ flowing through it and voltage difference $v_b(t)$ across it. We can consider the circuit as having no voltage, no current, and no external bias or driving fields at $t = -\infty$, and then adiabatically turn on these fields. In this picture it is possible to transform from branch current and voltage to branch charge and flux, which are required for a Hamiltonian description:
\begin{align}
    Q_b(t) &= \int_{-\infty}^t i_b(t^\prime) dt^\prime \\
    \Phi_b(t) &= \int_{-\infty}^t v_b(t^\prime) dt^\prime
\end{align}
As described in \cref{sec:SCQ}, we can quantize a circuit by writing its Hamiltonian in terms of charge and flux operators defined at each node. It is therefore required to transform from branch coordinates to node operators. This can be a complex procedure, which is well described in refs. \cite{voolIntroductionQuantumElectromagnetic2017,kermanEfficientNumericalSimulation2020} (an alternative approach using loop fluxes is described in \cite{ulrichDualApproachCircuit2016}); here we give a brief summary. An excellent review with simple worked examples can be found in ref. \cite{rasmussenSuperconductingCircuitCompanionan2021}.

First, we must write a circuit graph so that any connected nodes always have a capacitive branch between them, as in \cref{fig:graph}. This is physically accurate, as any parts of a physical device that are close enough to couple inductively or through a junction will also have some stray capacitance. From this graph of $N$ nodes we define the 
$N\times N$ capacitance matrix $\mathbf{C}$ with off-diagonal elements $-C_{jk}$, the negative of the capacitance between nodes $j$ and $k$. The on-diagonal elements are defined so that each row and column sums to 0, and the matrix is symmetric by default.

We then define a spanning tree of capacitances, which connects every node to ground through a path of only capacitive couplings without forming any loops. This means that every node has a single unique path to ground through a series of capacitive branches. Summing the branch fluxes from ground to the $j$th node thus defines the node fluxes $\Phi_j$. 
Any external bias flux threading a loop is taken to add on branches which are not part of the tree. We can then use node fluxes as our coordinates and write the ``potential'' energy in terms of these coordinates. For instance, an inductor $L_{jk}$ connecting nodes $j$ and $k$ will contribute energy $E = (\Phi_j-\Phi_k + \Phi_{ext,jk})^2/2L_{jk}$, where $\Phi_{ext,jk}$ is the  bias flux assigned to this branch of the loop. Likewise, the energy contribution from a Josephson junction depends on the difference in node fluxes across it (plus any bias flux assigned to that part of a loop), $E = -E_J \cos \left[(\Phi_j-\Phi_k+\Phi_{ext,jk})/\varphi_0\right]$, where $\varphi_0 \equiv \hbar/2e$ is the reduced flux quantum.

Since we are using node fluxes as our coordinates, we can write the ``kinetic'' energy in terms of the node fluxes' time derivatives, which are equal to the classical node voltages. The energy contribution from a capacitor connecting nodes $j$ and $k$ is then $E = C_{jk}(\dot{\Phi}_j-\dot{\Phi}_k)^2/2$, i.e., it depends on the difference of these node flux time derivatives across it. Combining the kinetic and potential energies we can write the Lagrangian $\mathcal{L}$. From this we can define the node charges as the canonical momenta, $q_j \equiv \partial \mathcal{L} /\partial\dot{\Phi_j}$, and write the classical Hamiltonian in terms of node charges and fluxes, plus bias voltages and fluxes. At this point we change charge/flux coordinates to operators, change to dimensionless operators $\hat{n}_j \equiv \hat{q}_j/2e$, $\hat{\phi_j}\equiv \hat{\Phi}_j/\varphi_0$ and obtain the quantum Hamiltonian as in \cref{eq:coupled-transmon}. This procedure for defining node fluxes and charges ensures the canonical commutation relation $\left[\hat{\phi}_j,\hat{n}_j\right]=i$.

This process sounds daunting even when summarized. Luckily, the procedure is quite straightforward, if complicated, and has been both written about in great detail \cite{voolIntroductionQuantumElectromagnetic2017,kermanEfficientNumericalSimulation2020} and has been integrated into open-source codes like scqubits, SQCircuit, and CircuitQ \cite{groszkowskiScqubitsPythonPackage2021a,rajabzadehAnalysisArbitrarySuperconducting2023,aumannCircuitQOpensourceToolbox2022}. A user can input an arbitrary circuit, follow the procedure, and obtain a numerical solution of the eigenstates and energies. In practice, this requires truncating the charge and flux basis states so that the Hilbert space is finite and can be diagonalized numerically. However, for a circuit with any significant complexity, a complete numerical solution can be computationally infeasible even when bases are truncated. One solution is to perform \emph{hierarchical diagonalization} \cite{kermanEfficientNumericalSimulation2020,chittaComputeraidedQuantizationNumerical2022,rajabzadehAnalysisArbitrarySuperconducting2023}. In this approach, a designer uses their knowledge of a circuit to divide it into subcircuits, each of which has well-defined eigenstates that can be easily solved. The subcircuits are described using a basis of these eigenstates, which again is truncated at some finite dimension. The subcircuits are then coupled together and solved using the product space of these subcircuit basis states. This is similar to a perturbative approach but differs in that it performs exact diagonalization within subcircuits, allowing efficient treatment of degenerate or near-degenerate states without relying on small parameter expansions.

For ease of computation it is highly advantageous to have a Hamiltonian that is \emph{mostly} diagonal---that is, a Hamiltonian with few nonzero off-diagonal matrix elements. A major speedup can be achieved by moving from charge and flux variables based on the circuit nodes to charge and flux variables that can be quantized into an approximately diagonal Hamiltonian for a subcircuit. This is analogous to transforming to coordinates that represent normal modes of oscillation in a classical system. Rigorous recipes for this procedure have been developed and are implemented in open-source codes SQCircuit and scqubits \cite{kermanEfficientNumericalSimulation2020,chittaComputeraidedQuantizationNumerical2022,rajabzadehAnalysisArbitrarySuperconducting2023}.

In addition to the challenge of solving an arbitrary circuit, it is typically desirable to have an effective quantized Hamiltonian written in terms of oscillator creation/annihilation operators and Pauli matrices, as in \cref{eq:cQED}. Such an effective Hamiltonian is more amenable to analytical theory and is usually more useful for quantum gate design and error correction architecture design. One solution is to use knowledge of the intended design to build up subcircuits that are weakly coupled---for instance, the coupled transmon and oscillator in \cref{fig:cQED} can be solved as an arbitrary circuit, but it is easier to recognize the transmon and oscillator modes and then add in the coupling in terms of these mode operators. Another approach is to exactly diagonalize the Hamiltonian, then write it in its eigenbasis with a set of Kerr, cross-Kerr, and higher-order terms as in \cref{eq:diagKerr}. This approach has the advantage of directly giving the energy spectrum, but is less useful if environmental conditions (flux or charge biases or drive tones) will be changed, as these can completely change the mode structure. Methods for rapidly calculating terms of interest such as inter-qubit couplings and dispersive couplings have been developed \cite{solgunSimpleImpedanceResponse2019a,solgunDirectCalculation$ZZ$2022}, making the procedure less computational intensive.

It is possible to connect nonlinear circuit elements in such a way that a Lagrangian cannot be defined via this procedure. For instance, a Josephson junction in parallel with a \emph{quantum phase slip junction} has a Hamiltonian description, but deriving a consistent Lagrangian is nontrivial due to the nonlocal behavior. A phase slip junction is the dual of a Josephson junction, with Hamiltonian contribution $H_Q = -E_Q \cos\left(2\pi\frac{q}{2e}\right)$---the appearance of the charge in the nonlinear term breaks the standard procedure \cite{osborneSymplecticGeometryCircuit2024}. Likewise, the Lagrangian for a non-reciprocal device cannot be straightforwardly written down since the branch flux would depend on the direction we choose to calculate it \cite{labarcaToolboxNonreciprocalDispersive2024}. Fortunately, frameworks for dealing with these situations have been formulated in geometric \cite{egusquizaAlgebraicCanonicalQuantization2022,osborneSymplecticGeometryCircuit2024,parra-rodriguezGeometricalDescriptionFaddeevJackiw2024} and exactly quantized \cite{parra-rodriguezExactQuantizationNonreciprocal2025} terms as well as including open-systems effects \cite{labarcaToolboxNonreciprocalDispersive2024}. We refer the reader to Ref. \cite{parra-rodriguezGeometricalDescriptionFaddeevJackiw2024} for an excellent summary of the literature. 

\subsection{Open-systems effects}\label{sec:purcell}
The Hamiltonian description of a circuit explained above does not include lossy effects. Here one may use a master equation or non-Hermitian Hamiltonian \cite{zhouRapidUnconditionalParametric2021} to describe the circuit. However, in the most commonly-considered case, \textit{Purcell decay}, a full quantum treatement is unnecessary. Purcell decay refers to the decay of a qubit via its coupling to a readout resonator, which itself is coupled to the lossy external microwave environment \cite{krantzQuantumEngineersGuide2019}. The term is sometimes also used to refer to decay to the external microwave environment due to any intended couplings, e.g., from qubit drive lines. In the case where the qubit and readout resonator are far detuned ($\Delta \equiv \omega_R-\omega_Q\gg g$) and the resonator has a high quality factor ($Q\equiv \omega_R/\kappa \gg 1$, where $\kappa$ is the resonator loss rate), the Purcell-induced decay rate is
\begin{equation}\label{eq:purcell}
    \Gamma_1^\mathrm{Purcell} = \frac{g^2}{\Delta^2}\kappa
\end{equation}
. This expression holds for a single resonator in the dispersive regime $g\ll\Delta$. When the resonator is driven with a coherent state with mean photon number $\bar{n}$ and the output is amplified with a phase-preserving amplifier, qubit state information is acquired at a rate
\begin{equation}\label{eq:measurementrate}
\Gamma_\mathrm{meas} = \frac{\chi^2}{\chi^2+\kappa^2}\kappa\bar{n}
\end{equation}
\cite{krantzQuantumEngineersGuide2019}. We see here a tension: the measurement rate increases with $\kappa$ and $\chi$, but so does the Purcell decay rate (since $\chi\sim g^2$). The solution is to add a \textit{Purcell filter} \cite{bronnReducingSpontaneousEmission2015}, which reduces the effective loss rate at the qubit frequency. Such a filter may be a bandpass near the resonator frequency or a bandstop near the qubit freuqency. This replaces $\kappa$ with $\kappa_\mathrm{eff}(\omega_Q)$ in \cref{eq:purcell}. Here $\kappa_\mathrm{eff}$ can be calculated as the linewidth of a resonator at $\omega_R$ that would produce the same real admittance as that seen by the qubit at $\omega_Q$. In the case of a bandpass filter at frequency $\omega_F\approx \omega_R$ with bandwidth $\kappa_F$, this yields
\begin{equation}\label{eq:Purcellfiltered}
\Gamma_1^\mathrm{Purcell} = \frac{g^2}{\Delta^2}\frac{\omega_Q}{\omega_R}\frac{\kappa_F}{2\Delta}\kappa
\end{equation}
where we have taken the limit that the filter bandwidth is much larger than the detuning between filter and resonator \cite{seteQuantumTheoryBandpass2015}. 

Note that this expression is exactly what one would obtain by treating the qubit as a classical harmonic oscillator and calculating $\Gamma_1 = \mathrm{Re}\left\{Y(\omega_Q)\right\}/C_Q$, where $Y$ is the admittance seen by the qubit and $C_Q$ is the total qubit capacitance. This classical admittance approach provides a convenient shortcut for transmon qubits which are only weakly anharmonic, and can function for any Purcell filter geometry. It also works for non-resonant couplings such as those to drive lines, and allows one to easily calculate the impact of the interference of qubit radiation with iteslf at different physical locations \cite{yenInterferometricPurcellSuppression2025,patelWavesinspacePurcellEffect2025}. For more anharmonic circuits circuits a prefactor $\beta \equiv \bra{0}\hat{n}\ket{1} / n_{01}^{QHO} $, i.e., the charge number matrix element at the node where the coupling occurs normalized by the harmonic oscillator charge matrix element, is required to account for the different effect of charge coupling on the circuit states. This is for the case where the qubit is capacitively coupled to the lossy mode or line; for inductive coupling we replace $\hat{n}\rightarrow\hat{\phi}$. Note that both these matrix elements are 1 for a harmonic oscillator circuit and near 1 for a transmon qubit.

Purcell decay is best addressed at the circuit graph level, as it is intrinsic to the couplings that are built into the graph and can be addressed as a microwave filtering problem. However, as we discuss in \cref{sec:EPR}, it can also be directly simulated from the device layout.

\subsection{From Layout to Graph}

The task of mapping a physical device layout to a circuit graph requires first defining features that can be considered as the branches and nodes of the circuit graph. For instance, consider the device shown in \cref{fig:layoutToGraph} \cite{mauryaDemandDrivenDissipation2024}. We treat the entire ground plane as a single node, treat each interdigitated coupling capacitor as a capacitive branch, treat each ``cross'' feature as one node of a capacitive branch and the ground plane as the other node, treat bond pads as 50 $\Omega$ resistors, etc. Care must be taken to account for the finite size of each feature, which manifests as self-inductance of capacitor pads and self-capacitance of inductive traces. Distributed-element resonators can be turned into lumped effective models \cite{besedinQualityFactorTransmission2018,minev2021circuitquantumelectrodynamicscqed}, with the caveat that this transformation might require frequency-dependent effective capacitances and inductances.

There is no rigorous standardized protocol for making these decisions. Device designers use intuition and experience to determine which components can be treated as lumped elements and which require modeling as distributed structures, which couplings are likely to be significant and which can be ignored, and which nonlinearities are significant and must be included (e.g., spurious Josephson junctions). Luckily, it is straightforward to \emph{check} if a circuit graph is accurate enough once it is defined and the device has been simulated. As described in \cref{sec:simulation}, a simulation can give the inductance and capacitance matrices for the circuit graph, but it can also directly given the diagonalized energy levels of the effective Hamiltonian. Solving the circuit Hamiltonian and then comparing to the directly-simulated energies can help a designer determine if the circuit graph requires revision. Care must be taken to determine whether discrepancies between the two approaches are due to improper definition of the circuit graph or inaccurate simulation of the eigenenergies and/or circuit element parameters---see \cref{sec:simAccuracy}.

\section{Device Layout} \label{sec:layout}

In the end, a device designer must produce a physical design layout that can be fabricated. This is perhaps the most challenging step of the design process. Given a fabrication process, the layout determines the effective circuit graph, including undesired stray couplings, and the values of each of the capacitive, inductive, and Josephson branch elements. It also determines couplings to lossy elements and drive lines, setting limits on coherence and gate speed. Here we discuss several concerns that go into creating a layout.

\subsection{Circuit Parameters}
The first priority in layout design is to create a device with the correct circuit parameters--capacitances, inductances, transmission line resonances, distributed couplings, and Josephson energies. Josephson energy is the most straightforward to design for---one simply designs a Josephson junction with a junction area that gives the correct critical current, given the critical current density of a fabrication process. Edge effects in junctions may mean that the critical current does not precisely scale with area for small junctions (empirically, less than $\sim$ 200 nm on a side), but these can generally be accounted for phenomenologically once a fabrication process is well-characterized.

Capacitance between co-planar structures cannot be easily computed analytically, nor can inductance of any but the simplest structures, nor can distributed coupling strength. These can be simulated quite straightforwardly, as described in \cref{sec:simulation}. A designer can determine capacitance and inductance values after making a design and conducting a (computationally-intensive) simulation for these parameters. Distributed-element behavior can also be simulated, although typically this requires a different solver (see \cref{sec:simulation}). Simulations can be time-consuming, so it is preferable to minimize the number of times a design must be modified and re-simulated. Fortunately, the \emph{scaling} of most parameters can be easily approximated, at least over a small range of variation. Capacitance between two features roughly scales linearly with the area of the smaller feature (i.e., the overlap area) and the inverse of the center-to-center distance \cite{Shanto2024squaddsvalidated}. Inductances roughly scale linearly with the length; mutual inductance roughly scales with the overlap length and the inverse of center-to-center distance. Resonant frequencies of transmission-line resonators scale roughly linearly with length, provided one takes into account any inductive or capacitive loading with a lumped-element model.

One must also take into account any kinetic inductance expected in the superconducting film. This typically manifests as an additional inductance per square---that is, for a trace of width $w$ and length $l$, the inductance scales as $l/w$. The inductance per square depends on details of fabrication such as film material, film thickness, and growth conditions. Kinetic inductance can be accounted for in simulations (see \cref{sec:KI}) but requires careful modification of standard simulation modalities.

In practice, a device is almost never designed from scratch. A designer takes components from previous devices and combines them to form a new design, adjusts the component sizes and arrangement based on their best guess of the correct scaling, then simulates. Based on this simulation the designer refines the geometry based on quasi-linear scaling, simulates again, and repeats until parameters are within tolerance.

\subsection{Crosstalk}\label{sec:crosstalk}
Crosstalk, generically, refers to any unwanted change in the behavior of one part of a device (e.g., a qubit) due to some operation on some other part of the device (e.g., a gate on a nearby qubit). \emph{Classical} crosstalk, also called microwave crosstalk, refers to unwanted stray couplings of electromagnetic fields between different circuit elements \cite{abramsMethodsMeasuringMagnetic2019,daiCalibrationFluxCrosstalk2021, wangControlMitigationMicrowave2022,kettererCharacterizingCrosstalkSuperconducting2023,yangFastUniversalScheme2024}. For example, a drive line for one qubit can also drive nearby qubits, causing unwanted rotations of those qubit states. This may be modeled as an undesired branch between two nodes in a circuit graph. As discussed in \cref{sec:graph}, in principle any two areas on a circuit couple capacitively and inductively. In practice, these couplings can be ignored if they are below some level. When they are significant, couplings cause crosstalk between different circuit elements. Classical crosstalk is a problem of improperly translating a desired circuit graph to a layout, and so must be addressed by changing the layout or calibrating the crosstalk into control schemes. The latter can be challenging: calibration of crosstalk between $n$ subsystems is a problem which scales as $n^2$ for linear crosstalk and can grow combinatorially (up to $n!$) for strong, high-order nonlinear interactions. Luckily, classical crosstalk is usually represented well by a sparse matrix \cite{daiCalibrationFluxCrosstalk2021,winickSimulatingMitigatingCrosstalk2021,kettererCharacterizingCrosstalkSuperconducting2023,yangFastUniversalScheme2024}.

Layout design can address classical crosstalk in several ways. The simplest is to move elements further apart, reducing coupling. This is effective, but adds bulk to the circuit and can make it difficult to also have desired couplings. Another approach is to use some aspect of the geometry to suppress coupling of nearby elements. For instance, two co-planar waveguide (CPW) transmission lines that cross each other at right angles without touching will nominally have 0 coupling.
Finally, the field from an element may be confined by shielding it with some metal. This is one purpose of air-bridge crossovers, which connect two metal sections while bridging over anything in between \cite{abuwasibFabricationLargeDimension2013,chenFabricationCharacterizationAluminum2014,dunsworthMethodBuildingLow2018,sunFabricationAirbridgesGradient2022,janzenAluminumAirBridges2022,taoFabricationCharacterizationLow2024,buTantalumAirbridgesScalable2024}. These crossovers are quite effective at confining field, as they can transform a CPW (typically used for routing signals and for distributed-element resonators) into something more like a coaxial cable \cite{abuwasibFabricationLargeDimension2013}. It is therefore standard practice to place them everywhere on the chip possible. Unfortunately crossovers require several steps of fabrication and so may be lossy (see below), limiting their placement near qubits. Another approach to field confinement is to use a ``flip-chip'' geometry, where two substrates are bonded together with a $\sim 1-10 \ \mu$m gap between them \cite{rosenberg3DIntegratedSuperconducting2017}. The device is then laid out in two planes on the faces of these substrates. This geometry makes it easier to bring coupling structures close together with vacuum in between, allowing them to be smaller \cite{liVacuumgapTransmonQubits2021}. More importantly, it confines field by providing more nearby metal for field lines to terminate in \cite{kosenSignalCrosstalkFlipChip2024}. Both these effects suppress long-range couplings and reduce classical crosstalk. 

Importantly, any attempt to reduce classical crosstalk will affect circuit parameters---it is impossible to modify the geometry in such a way that it \emph{only} reduces some stray coupling. It is crucial to include crosstalk suppression in the design process from the beginning, optimizing it in parallel with changing geometries to match target circuit parameter values.

\begin{figure}
\includegraphics[width=3.3in]{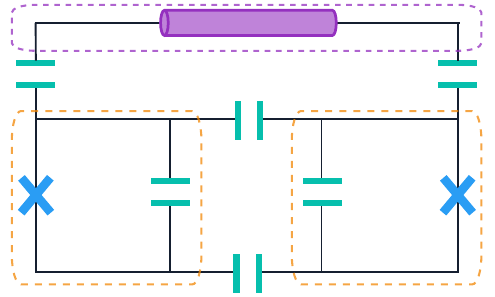}
\caption{\label{fig:multipath} Circuit diagram of a typical multi-path coupling between two transmon qubits (within the dashed orange boxes). Combining a direct capacitive coupling with an intermediate coupling resonator (purple) allows the designer to cancel the effective $ZZ$ interaction induced by the transmons' higher excited states.}
\end{figure}

\emph{Quantum} crosstalk refers to any unwanted coupling between two quantum degrees of freedom that is a byproduct of an intended coupling. Essentially, it is a failure to translate a desired effective Hamiltonian to a circuit graph, even if that circuit graph is then perfectly implemented in a layout. It is therefore a problem that should be addressed at the circuit graph design level when possible. For example, transmons which are capacitively coupled to allow for interactions in the qubit space (the lowest 2 energy levels) also have interactions with higher levels. The dominant effect of these higher level interactions is ``ZZ crosstalk'', an unwanted shift of the energy of the $\ket{11}$ state \cite{sheldonProcedureSystematicallyTuning2016}. This crosstalk can be addressed with quantum control \cite{mitchellHardwareEfficientMicrowaveActivatedTunable2021,weiHamiltonianEngineeringMulticolor2022,tripathiSuppressionCrosstalkSuperconducting2022}. However, it is often preferable to suppress it as much as possible with design. This can be done by using multiple coupling paths (see \cref{fig:multipath}), where two transmons are directly coupled and also both couple to an intermediate system (usually a linear resonator or a tunable transmon) \cite{yanTunableCouplingScheme2018,mundadaSuppressionQubitCrosstalk2019,kandalaDemonstrationHighFidelityCnot2021}. The multiple couplings allow fine-tuning of the Hamiltonian and suppression of ZZ crosstalk while maintaining the desired coupling. Again, such a modification to the device design must be included from the beginning so that circuit parameters can be correctly matched.

\subsection{Connectivity}\label{sec:connectivity}
While crosstalk---unwanted couplings---must be suppressed, a device layout must still produce the \emph{desired} couplings between subcircuits. Consider a subcircuit consisting of a tunable transmon qubit in a large processor architecture. The transmon must be capacitively coupled to a charge drive line; inductively coupled to a flux drive line; capacitively coupled to each desired nearby transmon, tunable coupler, bus coupling resonator, etc; and capacitively coupled to its readout resonator. However, the transmon must not couple to other drive lines, resonators, or couplers. This quickly becomes an issue as the number of subcircuits grows---how does one route a signal from one part of a chip to another without coupling to anything in between? As discussed in \cref{sec:crosstalk}, crosstalk suppression can be achieved by routing signals on CPW lines that use airbridge crossovers to cross each other at right angles; by using airbridge crossovers to confine field under the crossovers; and by using a flip-chip geometry to further confine field. A flip chip structure also allows designers another degree of freedom, as elements can be put on separate layers to reduce crowding, with indium bump contacts between the chips to form superconducting links \cite{rosenberg3DIntegratedSuperconducting2017,foxenQubitCompatibleSuperconducting2017}. Recent work has focused on extending to full 3D integration, using superconducting through-silicon vias (TSVs) to connect layers on opposite sides of a silicon substrate \cite{vahidpourSuperconductingSiliconVias2017,yostSolidstateQubitsIntegrated2020a,rosenbergSolidStateQubits3D2020a}. This allows much more flexibility in device design, although the technology has yet to see wide use publicly.

Another consideration is the \emph{capacitance budget} of a subcircuit. Each capacitive coupling between a subcircuit and its neighbors adds in parallel to the subcircuit capacitance(s), which must be reduced to compensate and maintain a constant total. In some architectures, the coupling capacitances eventually dominate the total capacitance, and it becomes impossible to increase a coupling without weakening all others. This is especially a concern in subcircuits such as fluxonium qubits, where the charge coupling matrix elements are quite weak \cite{manucharyanFluxoniumSingleCooperPair2009} and so coupling capacitances must be a significant fraction of the overall capacitance \cite{moskalenkoHighFidelityTwoqubit2022,dingHighFidelityFrequencyFlexibleTwoQubit2023}. A similar problem can occur with inductive couplings. One alternative is to use \emph{galvanic} coupling, where two subcircuits share a linear or nonlinear inductive element (i.e., an inductor or Josephson element). By using a SQUID as the galvanic coupler, it is possible to tune the coupling rapidly in-situ \cite{weissFastHighFidelityGates2022,zhangTunableInductiveCoupler2024}, and the galvanic coupling budget is generally large enough to allow for coupling to at least 4 other qubits (although complexity may make this approach undesirable). 

These concerns exist in the context of quantum error correction (QEC) architecture design. In general, the more connected the qubit graph, the easier QEC is to accomplish. That is, if each qubit can be made to run high-fidelity two-qubit gates with a greater number of other qubits, then fault-tolerant QEC can be achieved with fewer resources (qubits and operations) and at a higher gate error rate \cite{gottesmanFaultTolerantQuantumComputation2014}. At some point it becomes impractical to add more fixed couplings. Directly coupling two distant qubits together then requires some long-range controllable interaction such as controlled transmission of microwave photons \cite{kurpiersDeterministicQuantumState2018,leungDeterministicBidirectionalCommunication2019,kannanDemandDirectionalMicrowave2023,almanaklyDeterministicRemoteEntanglement2025}, resonant couplings between modular elements \cite{zhouRealizingAllallCouplings2023}, or measurement-based remote entanglement \cite{rochObservationMeasurementinducedEntanglement2014,kimchi-schwartzStabilizingEntanglementSymmetrySelective2016,greenfieldStabilizingTwoqubitEntanglement2024}.

\subsection{Interface Losses}\label{sec:interfaces}
It has now been well established that interface layers between metal (superconductor) and substrate, between substrate and air (or vacuum), and between metal and air (or vacuum) can all act as dielectrics that are many orders of magnitude more lossy than crystalline silicon/sapphire  dielectric substrates \cite{gaoExperimentalEvidenceSurface2008,oliverMaterialsSuperconductingQuantum2013a}. Microwave loss tangents of substrates are of order $\tan\delta \sim 10^{-6}-10^{-7}$ while interfaces can be as lossy as $\tan\delta \sim 10^{-2}-10^{-3}$. The decay rate $\Gamma$ of a qubit as limited by dielectric losses is 
\begin{equation}
    \Gamma = \beta \omega_q \sum_i P_i \tan \delta_i
    \label{eq:loss}
\end{equation}
where $\omega_q$ is the qubit frequency, $P_i\in [0,1]$ is the participation ratio of the ith dielectric material (the fraction of total electric field energy that is in this material),  $\tan \delta_i$ is the microwave loss tangent of that material, and $\beta$ is a dimensionless prefactor indicating the strength of the charge decay matrix element. For a harmonic oscillator $\beta = 1$ and for a transmon $\beta \approx 1$. This model assumes low-field, linear dielectric response, valid at millikelvin temperatures typical of superconducting qubit operation.

\begin{figure}
    \centering
    \includegraphics[width=3.3in]{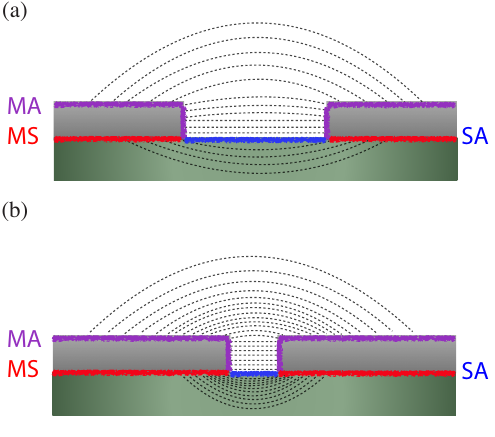}  
    \caption{Qualitative depiction of the electric field distribution between two capacitor plates that are close together (a) or far apart (b). For the same total field energy, the close-together plates concentrate more field in surface and interface layers. The metal-air (MA) interface is shown in purple, the metal-substrate (MS) interface in red, and the substrate-air (SA) in blue; interfaces are drawn as rough boundaries to emphasize their amorphous nature. These interfaces are typically much lossier than the substrate (green) or vacuum, with microwave loss tangents 3 to 5 orders of magnitude higher than bulk silicon or sapphire. This increased field participation in lossy interfaces leads to higher dielectric loss, as described by \cref{eq:loss}}
    \label{fig:interface_loss}
\end{figure}

Much of the progress in the increasing qubit energy relaxation times ($T_1$) can be attributed to reducing the lossiness of these interfaces---reducing $\tan \delta_i$, a \emph{fabrication} challenge \cite{siddiqiEngineeringHighcoherenceSuperconducting2021b}---and to designing qubits that reduce the participation ratio of these layers---reducing $P_i$, a \emph{design} challenge. The general approach to reducing interface participation ratio is to move capacitive elements farther from each other while increasing their area. This keeps the capacitance constant while decreasing the concentration of electric field energy in interface layers. See \cref{fig:interface_loss}. Increasing element size almost uniformly reduces dielectric loss \cite{wangSurfaceParticipationDielectric2015}. This was the principle behind the development of the 3D transmon \cite{paikObservationHighCoherence2011}, which at the time of its invention was 1-2 orders of magnitude more coherent than typical small-element planar transmons. The tradeoff is that this increased size in turn means larger stray couplings at larger distances, making it harder to suppress crosstalk in a large planar circuit. Making circuit elements larger also adds bulk, which can be a significant consideration when more than a few qubits are placed on a chip.

Another method to reduce interface losses is to design the geometric features with smaller perimeters, fewer sharp corners, and fewer narrow sections. All of these changes reduce the participation of lossy surface layers \cite{martinisSurfaceLossCalculations2022a}. Similarly, metal films incorporating oxide layers (for junction fabrication) may be lossier than single-layer films \cite{smirnovWiringSurfaceLoss2024}. These films are typically only present in the connection between Josephson elements and larger-scale features; reducing the lengths of these connections can reduce the loss. 

\subsection{Radiation}\label{sec:radiation}
A circuit may couple to the external electromagnetic field via antenna modes. Devices that are meant to be embedded in 3D cavities are typically designed with large (mm-scale) antenna structures to provide dipole coupling to one or more cavity modes. This coupling can be modeled several ways, as discussed in \cref{sec:em-qm}. Typical design rules for microwave antennae apply, including matching the antenna length to the wavelength (or some integer divisor), loading the ends with capacitive structures, and avoiding spurious antenna resonances at unwanted frequencies.

In planar circuit devices, radiation is almost always an undesired loss mechanism. Small feature sizes and robust waveguide designs typically do an excellent job suppressing radiation in the microwave regime. However,  qubit geometries can make excellent antennae for mm-wave radiation \cite{raffertySpuriousAntennaModes2021a}. This can lead to resonant absorption of pair-breaking radiation \cite{liuQuasiparticlePoisoningSuperconducting2024}, generating quasiparticles  and inducing them to tunnel across Josephson junctions \cite{houzetPhotonAssistedChargeParityJumps2019} or trap in internal junction Andreev bound states \cite{haysDirectMicrowaveMeasurement2018,farmerContinuousRealtimeDetection2021,elfekyQuasiparticleDynamicsEpitaxial2023a}. This radiation can reach the device either by propagating through free space (if the device is not properly enclosed, see \cref{sec:enclosure}) \cite{barendsMinimizingQuasiparticleGeneration2011a} or by propagating down microwave lines \cite{wangMeasurementControlQuasiparticle2014}. Care must be taken to ensure that the qubit structure does not act as a resonant antenna at frequencies that are poorly-shielded or poorly-filtered. In general it is preferable to push the antenna resonances to higher frequency, as it is easier to attenuate the higher-frequency radiation. This means making capacitive structures smaller and closer together, which is in tension with the need to minimize coupling to lossy interfaces discussed in \cref{sec:interfaces}.

\subsection{Enclosure and embedding}\label{sec:enclosure}
Here we briefly discuss the requirements for enclosing and embedding a superconducting device. For a comprehensive review of enclosure and embedding designs, see \cite{huangMicrowavePackageDesign2021}.

A superconducting circuit must be enclosed in a conductive cavity to prevent radiative coupling of both microwave and mm-wave radiation, as discussed above. This cavity will have its own resonant modes. When the circuit is designed to couple only to other circuits in plane (or on an adjacent plane, as in a flip-chip geometry), any coupling to cavity modes will be a potential source of loss or crosstalk. The typical approach is to make the cavity dimensions as small as possible and thus push mode frequencies higher, but this becomes infeasible as the chip size grows. Flip-chip geometries with indium bump bonds can make an effective cavity between the chips, greatly reducing volume and increasing mode frequencies. 

Off-resonant coupling is also a potential problem. Cavities are often made of normal metals (e.g., gold-plated copper) to aid with thermalization and to allow external flux biasing. In this case radiative coupling to the cavity walls can be a source of loss even if no energy enters a cavity mode, simply because conduction in the cavity walls is lossy. Solutions include moving the cavity walls further away from the chip \cite{huangMicrowavePackageDesign2021}, at the expense of lower cavity mode frequencies; making the cavity from a superconducting material or coating the backside of the chip with superconductor, at the expense of thermalization, flux biasing, and quasiparticle poisoning suppression \cite{iaiaPhononDownconversionSuppress2022}; and changing the on-chip design to confine the field more closely, at the expense of increased coupling to lossy interfaces.

In the case where the cavity is being used for bosonic qubit encoding, the cavity modes are deliberately designed to be lower frequency, with at least one mode engineered to have an ultra-high quality factor $Q$ and thus long-lived coherence. Usually, one or a few qubits are coupled to the high-coherence mode(s) and used to control it (them). Each qubit may also be coupled to a low- or moderate-$Q$ cavity mode to perform readout, or to a planar resonator for the same purpose \cite{axlineArchitectureIntegratingPlanar2016}. Microwave connectors with conducting pins that protrude into a cavity can act as antennae, coupling in charge drive signals and coupling out readout signals. Flux biases and flux drive tones are more difficult to bring into the cavity, as the physical structure is typically superconducting (to enable high $Q$) and thus expels magnetic field. Recent work has focused on this problem, either by designing special ``magnetic hose'' structures into the cavity \cite{gargiuloFastFluxControl2021} or by carefully designing the cavity to allow partial transmission of magnetic flux \cite{reshitnyk3DMicrowaveCavity2016,stammeierApplyingElectricMagnetic2018}.

Signals must also be brought to larger planar devices. In this case the typical approach is to use solder connections or wirebonds to connect a microwave coaxial connector to a microwave circuit board. This board uses normal-metal waveguide traces to bring the signals close to the chip, where wirebonds connect the waveguides to on-chip lines. Again the embedding board must be co-designed with the chip and enclosure to reduce radiative coupling to lossy elements. Connectivity can be a challenge, as typical devices only have wirebond pads available around the edges. 3D integration can solve this issue by using a fabricated interposer chip between the embedding board and the device, with the interposer routing signals through lines that are more compact and lower-loss than the embedding board, but without the strict coherence requirements on the quantum device \cite{rosenberg3DIntegratedSuperconducting2017}. To the best of our knowledge this has not been demonstrated publicly, but the recent devlopment of TSVs makes it a possibility.

\section{Electromagnetic Simulation} \label{sec:simulation}

Electromagnetic (EM) simulations are essential in the design process of superconducting quantum circuits, providing detailed insights into the circuit's electromagnetic behavior. These simulations help verify that the device layout accurately reflects the desired Hamiltonian description while also identifying potential issues, such as unwanted modes or parasitic couplings, that could degrade performance. Through this process, one can optimize the design layout before fabrication, reducing the need for costly fabrication iterations - significantly saving time and resources.

\subsection{Simulation tools} \label{sec:EMtools}
EM simulation tools solve Maxwell's equations numerically to provide detailed information about the electromagnetic fields and circuit parameters within the simulated device layout. These solvers can compute important quantities such as inductances, capacitances, impedance, and field distributions, all of which are key inputs for analyzing the electromagnetic properties of quantum circuits. Commercial finite element method (FEM) solvers like ANSYS HFSS/Q3D and COMSOL Multiphysics are the most commonly used in the community~\cite{Ansys2024, COMSOL2024, Shammah_2024}. They offer robust features, user-friendly interfaces, stable software, and comprehensive support, making them suitable for handling a wide range of complex simulations. These solvers are particularly useful for investigating coupling effects, wave propagation, and energy dissipation mechanisms in complex circuit layouts. Open-source alternatives such as Palace and ElmerFEM are also gaining traction~\cite{Shammah_2024, Malinen2013}, providing flexibility for customization, free access, OS-agnostic compatibility, and native high performance computing (HPC) features (in the case of Palace), which can be particularly advantageous for specialized quantum applications. 

Recognizing the unique requirements of superconducting quantum circuits, quantum-specific electronic design automation (EDA) tools have also started to emerge. For example, Keysight's \textit{EMPro} and \textit{Quantum Ckt Sim} ~\cite{Choi2023,naaman2024modelingfluxquantizingjosephsonjunction} provide a quantum device layout generation library and a tailored simulation environment, including advanced features such as flux quantization in superconducting loops, specifically designed to meet the needs of quantum circuit modeling.

\subsection{Finite-element simulations}\label{sec:FEM}
\begin{figure}
    \centering
    \includegraphics[width=\linewidth]{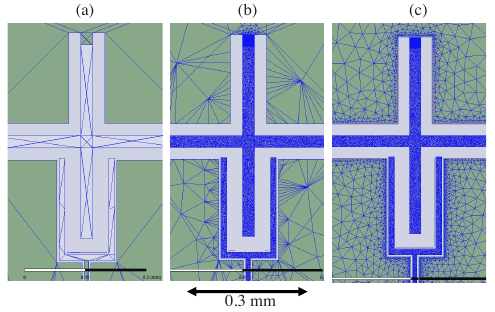} 
    \caption{(a) Typical finite-element mesh that is seeded at the start of a simulation. Here the simulation is of a transmon with a central cross capacitor, an outer ground plane, and a ``claw'' coupling capacitor. The Josephson junction is modeled as a lumped element inductor (small green square at top of cross). During the solving, each mesh element is treated as having a single value of electric and magnetic field (or a spatial dependence given by a low-order polynomial). When a mesh element is too large, this fails to capture significant field change within the element, leading to an inaccurate solution. (b) Mesh after several rounds of refinement. The refinement algorithm has correctly created a fine mesh on the transmon cross and coupling claw, but has not yet sufficiently refined the mesh on the ground plane. (c) Final mesh after solution convergence. The mesh elements are very small in any area where fields may be changing quickly, ranging from edge lengths of $0.1 - 1 \ \mu$m at the edges of the ground plane to below $0.1 \ \mu$m inside the qubit capacitor and coupling claw to below $0.01 \ \mu$m inside the junction.}
    \label{fig:mesh_convergence}
\end{figure}

FEM solvers function by first breaking space up into discrete elements. When expressed on this discrete \emph{mesh}, Maxwell's equations transform from differential equations to discrete difference equations. A solution to these difference equations on the mesh can be readily computed (although the computation may be time-consuming). Each mesh element is treated as if the field distribution across the element is either constant or changing with some low-order polynomial dependence on position (depending on solver settings) \cite{hfss_slide, fem_mesh_paper}. After a solution is computed, a new, finer mesh is generated based on the solver's best estimate of where the prior mesh was too coarse, as shown in \cref{fig:mesh_convergence}. A solver's mesh refinement algorithm is often proprietary, but a simple rule is to divide an element more finely if it has a large local field concentration or gradient. After mesh refinement a solution is computed again. The process is repeated until some simulated quantities of interest (such as mode frequencies or scattering parameters) change by less than a user-defined threshold from one iteration to the next. Depending on the solver, a user may also mandate multiple iterations below the convergence threshold, a minimum/maximum number of iterations, nonuniform convergence criteria, and other customizations.

Commercial software tools such as ANSYS use different back-end solvers for simulations aimed at extracting different quantities. A solution may give eigenmode frequencies and quality factors, showing the field distributions at those frequencies; it may give capacitance and inductance matrices between different geometric structures by solving electrostatic behavior; or it may give frequency-dependent scattering parameters. Each analysis requires a separate solver optimized for a specific computation. Choosing the right solver is essential for obtaining the correct quantities to analyze device behavior, as explained in \cref{sec:em-qm}.

\subsection{Kinetic inductance}\label{sec:KI}
While commercial simulation tools are capable of accurately simulating geometric inductance and other electromagnetic properties, they do not natively account for kinetic inductance $L_k$. Kinetic inductance arises in superconductors due to the inertial mass of mobile charge carriers. (Normal metals do not experience significant kinetic inductance since their charge transport is mostly diffusive---any inertial effects are swamped by rapid scattering.) This additional inductance beyond the geometric inductance becomes significant in quantum circuits operating at high frequencies or using superconducting materials with high disorder or low charge carrier density. To model kinetic inductance, we use the Mattis-Bardeen theory which describes the complex conductivity of superconductors at non-zero frequencies~\cite{Mattis1958}. For superconducting systems in the regime $T \ll T_c$ and $\hbar \omega \ll \Delta$ (where $T_c$ is the critical temperature and $\Delta$ is the superconducting gap), the kinetic inductance of a small wire at frequency $\omega$ can be expressed as:
\begin{align}
    L_k(\omega) = \frac{l}{w t \sigma_2 \omega } \approx  \frac{0.18 \cdot l \hbar \rho_n}{w t k_B T_c}
\end{align}
where $l$, $w$, and $t$ are the length, width, and thickness of the conductor, $\rho_n$ is the normal-state resistivity of the material, $\sigma_2 \approx \frac{k_B T_c}{0.18 \hbar \rho_n \omega}$ is the imaginary part of the complex conductivity, and $k_B$ is Bolztmann's constant \cite{Zmuidzinas2012, Hauer2024}. The key insights from this model are that materials which are poor normal conductors will have high superconducting-state kinetic inductance, and that this inductance is non-linear with respect to current. To incorporate this kinetic inductance into EM simulations, one must modify the material properties to reflect its complex conductivity. In COMSOL, for example, this is done by updating the conductivity to account for the frequency-dependent $\sigma_2$ term, which is proportional to $\alpha/\omega$, where $\alpha$ depends on $T_c$ and $\rho_n$. By including this imaginary component in the simulation, the solver can accurately model the kinetic inductance's contribution to device performance. This method can be also be applied in open-source tools such as Palace, allowing for precise simulations of quantum circuits that incorporate high-kinetic-inductance materials.

The solvers discussed above were designed for normal conductors; any superconducting effects must be tacked on. Recent work has focused on developing simulation techniques that are built up from first principles and incorporate superconductivity from the beginning \cite{phamSpectralTheoryNonlinear2024}. DEC-QED (discrete exterior calculus quantum electrodynamics) focuses in particular on 3D structures. By expressing the solution in terms of gauge-invariant fluxes, DEC-QED is able to accurately capture superconducting effects such as flux quantization, and is well-suited to modeling radiation and coupling effects \cite{phamFluxbasedThreedimensionalElectrodynamic2023}. Solvers incorporating these recent developments may provide significant advantages in accurate modeling of crosstalk, where the Meissner effect can significantly change couplings.

\subsection{Simulation accuracy}\label{sec:simAccuracy}
The accuracy of EM simulations is highly dependent on key simulation \emph{hyperparameters}, such as mesh density, convergence criteria, and solver tolerances. To achieve reliable results, we recommend setting the initial seed mesh element sizes so that there are at least 3 elements across any dimension of a metal feature or gap between features. For instance, a CPW trace with a width of $10 \ \mu$m and gap of $6 \ \mu$m should have an initial seed mesh with elements no longer than $3.3 \ \mu$m on the trace and $2 \ \mu$m in the gap. This rule ensures that the solver can recognize the difference between the edges and center of a feature, and therefore can easily recognize gradients between center and edges. Once the seed mesh is fine enough, the solver will automatically refine it further when needed. For Josephson junctions, a stricter seed mesh size of less than one-tenth of the junction dimensions is needed to capture field details essential for accurate energy participation analysis~\cite{Wenner2011, Minev2021, minev2021circuitquantumelectrodynamicscqed, Yuan2022, Shanto2024squaddsvalidated}. Of course it is always more accurate to use a finer seed mesh, but this accuracy must be balanced against the computational cost to generate and solve this finer mesh (i.e., the time required). The mesh sizes stated above seem to be a good compromise that ensures accuracy without being overly computationally intensive \cite{Shanto2024squaddsvalidated}.

Simulation convergence is determined by the change in some quantities (such as mode frequencies or capacitances) from one iteration to the next. A large change in these quantities indicates that mesh refinement is significantly changing the solution, and that further refinement would likely further improve accuracy. Once the change in, e.g., mode frequency between iterations is below some threshold, the convergence criteria are met and the simulation can end. The convergence threshold must be set to a low enough tolerance that the simulation accurately reproduces all quantities of interest, but high enough that the simulation finishes in a reasonable amount of time. Empirically we have found that a convergence threshold of less than $.05\%$ mean absolute change in mode frequency or capacitances between iterations is good enough for most purposes \cite{Shanto2024squaddsvalidated}, where the mean is taken across all mode frequencies (for eigenmode simulations) or all capacitance matrix elements (for electrostatic simulations).

\begin{figure}
    \includegraphics[width=3.3in]{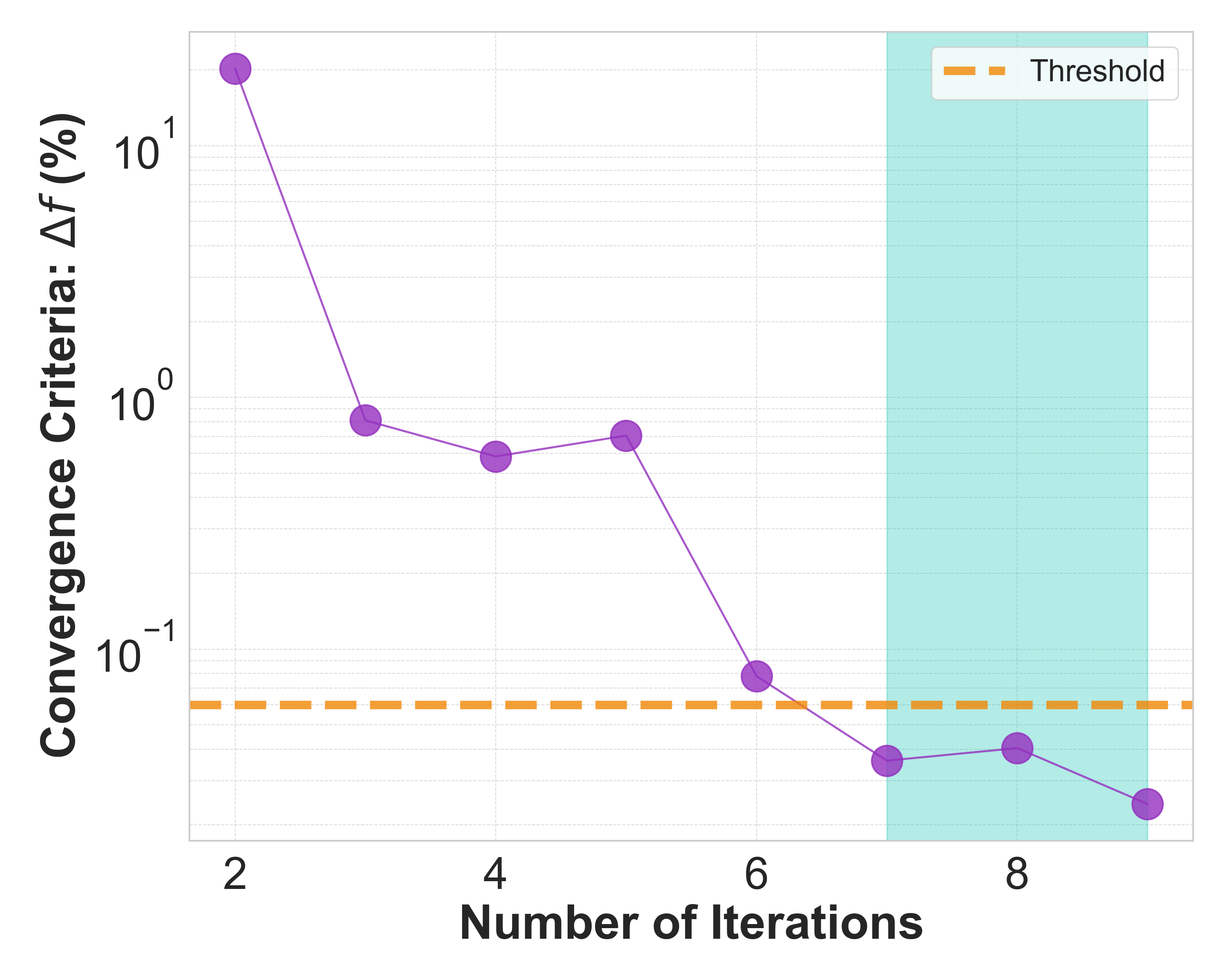}
    \caption{\label{fig:convergence} Example of a typical simulation convergence. Each iteration refines the solution, leading to a change in the solved mode frequencies. When this change is below some threshold for some number of consecutive iterations, the convergence criteria is met and the simulation is finished. Here we set the convergence criteria to be a mode frequency change less than $0.5 \%$ for 3 iterations in a row. The shaded region shows the iterations where the frequency convergence has been satisfied.}
\end{figure}

Sometimes the refinement of a solution from one iteration to the next does not appreciably change the convergence quantity, but does reveal new locations where the mesh is too coarse. This can result in a false convergence, where additional iterations (with finer meshing) would result in an appreciably different solution. See \cref{fig:convergence}. To avoid this issue, each simulation should undergo a minimum of one extra pass after the convergence threshold is met---when possible, two or three extra passes are preferred \cite{Shanto2024squaddsvalidated}. The tradeoff is, again, computational cost and speed, as these later iterations have the finest meshes and thus are the most time-consuming. 

\begin{figure}
    \centering
    \includegraphics[width=\linewidth]{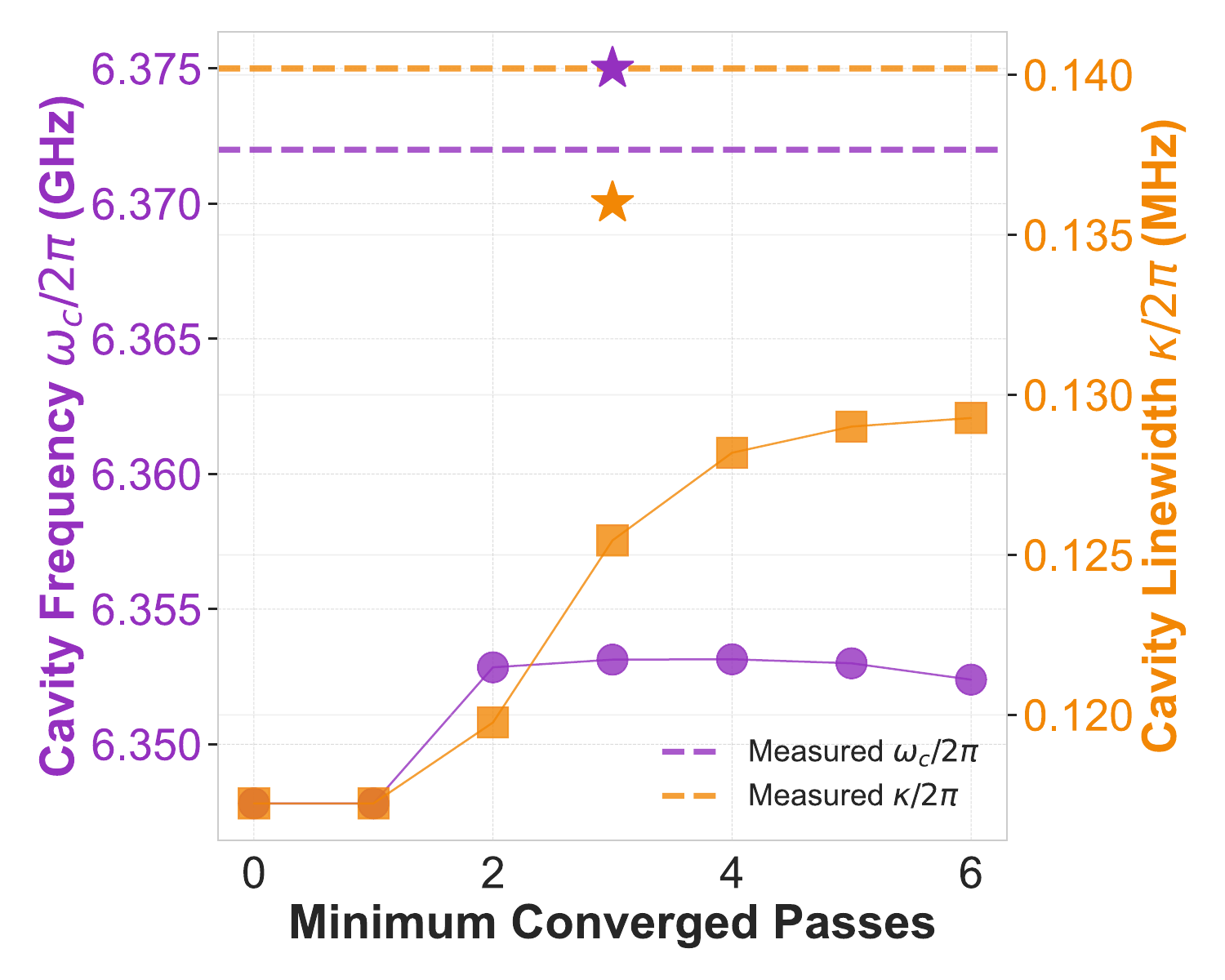}
    \caption{Demonstration of the dependence of simulation results on hyperparameters. Here we vary the number of consecutive passes that must satisfy the convergence threshold before the simulation can terminate. We compare the simulated resonant frequency (purple circles, left axis) and linewidth (orange squares, right axis) of a CPW resonator to the experimentally-measured values (dashed lines). In both cases the simulations saturate by 4 converged passes, but a small systematic error remains ($0.3\%$ for frequency and $8 \%$ for linewidth). The frequency error is mainly due to obfuscation of the fabrication etch bias, which are is a confidential parameter of the SQUILL foundry that fabricated the device. Including the etch bias reduces the error below 0.1\% (purple star marker). The linewidth error is improved below 3\% (orange star marker) when etch bias is included; we attribute the remaining error to impedance variations in the readout line, which simulations do not capture. }
    \label{fig:hyperparams}
\end{figure}

Another approach to improving accuracy is to \textit{add} convergence criteria, explicitly mandating that the change in one or more additional quantities is less than some threshold. For instance, as shown in \cref{fig:hyperparams}, mode frequency tends to converge faster than bandwidth. Mandating convergence on both may ensure an accurate solution without requiring as many extra passes after convergence. To the best of our knowledge there are no reported systematic studies of the advantages and drawbacks of using different convergence criteria, and we invite the community to rectify this!

\begin{figure}
    \includegraphics[width=3.3in]{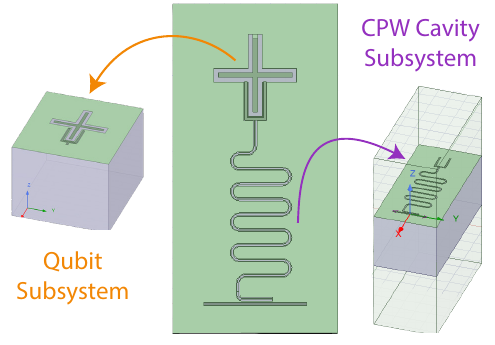}
    \caption{\label{fig:partition}
    Partitioning of a qubit-cavity system (center) into two overlapping sections. On the left, the qubit and coupling capacitor, with a stub to connect to the resonator, are simulated. On the right, the meandering CPW cavity with coupling capacitor and co-linear coupling to a feedline are simulated. The feedline is terminated with lumped $50 \ \Omega$ resistors (not shown) to represent the external microwave environment. We can use the results of both simulations to turn the subsystems into lumped-element models, then combine them to solve the Hamiltonian. }
\end{figure}

For complex layouts, it is generally better to simulate smaller sections instead of attempting to simulate the entire layout at once, which may be computationally prohibitive. For instance, on an in-house desktop workstation with an AMD Ryzen Threadripper PRO 3955WX 16-core 3.90 GHz processor and 128 GB of RAM, an accurate eigenmode simulation of a 2x2 mm area takes between minutes and hours, while a 10x10 mm area can take days. These smaller sections can then be turned into effective lumped circuit models \cite{minev2021circuitquantumelectrodynamicscqed} or effective quantized Hamiltonians and then combined. For example, for a design with a transmon capacitively coupled to a CPW cavity, one could simulate the transmon and coupling capacitor, then separately simulate the cavity and coupling capacitor. See \cref{fig:partition} for an illustration. This modular approach allows for the extraction of accurate circuit parameters within each section and makes the simulations manageable in terms of both time and computational resources. The results from these smaller simulations can later be combined during the quantum analysis stage to construct the full system’s Hamiltonian. This method is particularly useful for large-scale designs with repeated subsystems, where simulating the entire structure at once would be impractical and computationally expensive to achieve the necessary level of accuracy. However, this approach may not capture all important effects in layouts where components are closely spaced. In such cases, cross-talk or hybridization of electromagnetic modes between neighboring elements may not be fully accounted for, potentially limiting the accuracy of the final analysis. Therefore, careful consideration should be given when applying this method to designs with tightly packed elements or strong mode interactions. The designer should make the simulation sections large enough that couplings outside the edges are negligible, or overlap them to ensure that boundary effects do not dominate.

Balancing accuracy and computational resources is crucial in EM simulations. Higher accuracy, achieved through finer meshes and stricter solver settings, comes at the cost of increased simulation time and memory usage. To optimize this trade-off, it is highly beneficial to conduct convergence studies, refining mesh sizes and solver tolerances until further iterations result in no significant changes in Hamiltonian parameters. For an example of how the number of required consecutive passes below threshold can affect accuracy, see \cref{fig:hyperparams}. Additionally, properly defining boundary conditions (open, closed, or lossy), material properties (using cryogenic dielectric constants),  and simulation volumes, accounting for fabrication effects (see \cref{sec:fab}), and selecting the appropriate solver and methodology are critical for accurately modeling the physical environment of the device, as errors in these factors can compromise the reliability of the simulation results.

\subsection{From EM Simulations to Quantum Analysis} \label{sec:em-qm}
After completing EM simulations, the resulting data must be translated into quantum circuit parameters to construct the system's Hamiltonian. Typical solvers will provide either: inductance and capacitance matrices, along with field distributions for static fields; mode eigenfrequencies and quality factors, along with field distributions at those frequencies; or scattering parameters as 
a function of frequency, along with field distributions as a function of frequency. In each case the data can be used to calculate a Hamiltonian, although in practice the electrostatic and eigenmode simulations are preferred. Because the data provided are different, each simulation technique is better suited to a different analysis methods. Three common methods for performing this translation from simulation to Hamiltonian are the Energy Participation Ratio (EPR) method~\cite{Minev2021}, the Lumped-Oscillator Model (LOM)~\cite{minev2021circuitquantumelectrodynamicscqed}, and the Black-Box Quantization (BBQ) approach~\cite{PhysRevB.90.134504}. The EPR method uses eigenmode simulation data, LOM uses electrostatic data, and BBQ uses both. Here we give brief overviews of each method.

\subsubsection{Energy participation ratio analysis}\label{sec:EPR}
The EPR method \cite{Minev2021} is based on breaking the Hamiltonian into two parts: a linear part composed of oscillator modes and a nonlinear part consisting of the nonlinear inductive contributions from the Josephson elements.
\begin{align}
    \hat{H} &= \hat{H}_\mathrm{lin}+\hat{H}_\mathrm{nl} \\
    \hat{H}_\mathrm{lin} &= \sum_{m=1}^M \omega_m \hat{a}_m^\dag \hat{a}_m \\
    \hat{H}_\mathrm{nl} &= \sum_{n=1}^N -\left(E_{n}^{(J)}\cos \hat{\phi}_n + \frac{\hat{\phi}_n^2}{2}\right) \\
    \hat{\phi}_n &= \sum_{m=1}^M s_{mn}\sqrt{\frac{p_{mn} \omega_m}{2 E_n^{(J)}}}\left(\hat{a}_m+\hat{a}^\dag_m\right)
\end{align}
Here $\omega_m$ is the frequency of the $m$th mode, $M$ is the total number of modes considered, $\hat{a}_m$ is the lowering operator for the $m$th mode, $E_n^{(J)}$ is the Josephson energy of the $n$th junction, and $\hat{\phi}_n$ is the phase operator across that junction. The inductive energy participation ratio $p_{mn}$ is the fraction of the inductive energy in the $m$th mode that is stored in the $n$th junction (when only that mode is excited). The sign parameter $s_{mn} = \pm 1$ indicates which direction current flows across the $n$th junction in the $m$th mode relative to the first junction (one can pick $s_{m1} = 1$ arbitrarily). Both $p_{mn}$ and $s_{mn}$ can be directly computed from the FEM simulation field output, at which point the Hamiltonian can be solved numerically.

The EPR method's main advantages are its ease of use and its generality. It directly translates from eigenmode simulation results to an effective diagonalized Hamiltonian with mode frequencies, self-Kerr terms, and cross-Kerr terms (similar to \cref{eq:diagKerr}). There is thus a direct translation between a physical layout and the observed spectrum. The pyEPR package \cite{pyEPR} provides a simple Python API for running these calculations, and can be easily used by a designer who has no expertise in the physics behind the EPR method. EPR can capture an entire system's behavior at once, reducing the number of required simulations. It makes few assumptions and adapts well to complex circuits with multiple nonlinear elements, making it a versatile tool in quantum circuit design. EPR can also be used to compute the loss caused by dissipative elements. However, it can be computationally demanding for large systems, as it requires simulation of a subsystem large enough that all relevant junctions are included for each mode. Additionally, there is some evidence that EPR analysis can be more sensitive to simulation errors \cite{Yuan2022, Shanto2024squaddsvalidated}. This is likely because EPR depends on accurate simulation of the field strength in each Josephson element, which are typically the smallest features in the design. The simulation may finish based on the solver's convergence criteria while the junction fields are not yet accurately modeled. Modifying the solver to explicitly consider junction field in its convergence criteria may solve this issue (see \cref{sec:simAccuracy} for more discussion).

EPR analysis also allows easy calculation of the Purcell-limited decay rate of a mode. By terminating output ports with effective lumped-element resistors with resistance equal to the microwave environment impedance (i.e., $50 \ \Omega$), then calculating the field energy in these lossy elements, one can extract the effective real admittance and thus the Purcell decay rate. One could also simply use the quality factor of the qubit mode extracted from the eigenmode simulation itself. In either case a correction must be made if the qubit is highly anharmonic, as discussed in \cref{sec:purcell}

\subsubsection{Lumped oscillator model analyis}\label{sec:LOM}
LOM analysis uses the circuit graph representation of the device layout, treating it as a network of lumped capacitors, inductors, and junctions. Any distributed element (such as a CPW resonator) is approximated as an equivalent lumped circuit. An electrostatic simulation of the geometry then provides the capacitance and inductance matrices. Using the methods described in \cref{sec:graphtoH}, the resulting graph can be turned into a quantum Hamiltonian. 
Importantly, LOM analysis allows the physical layout to be divided into sections, each of which can be simulated independently to extract inductances and capacitances. Care must be taken to define these section boundaries so that there is no significant inductive or capacitive coupling across section boundaries, or to overlap sections to properly capture all couplings. Often it is useful to also define subsystems made of multiple sections. These subsystems are expected to implement a few, well-understood modes, and to be only weakly coupled to each other. The user can then solve and truncate the eigenstates of these subsystems, then couple them together and solve the whole Hamiltonian. LOM analysis is more flexible than other analysis techniques as it gives a circuit graph that can then be manipulated to test other possible designs. It requires only small electrostatic simulations that are computationally less intensive than full-device simulations. However, it may not accurately capture distributed effects or higher-order modes, potentially limiting its applicability for certain designs. Additionally, LOM analysis may require more detailed modeling of circuit graph connectivity and distributed effects to accurately extract the effective Hamiltonian, making it somewhat less plug-and-play compared to EPR analysis.

\subsubsection{Black-box quantization}
Lastly, the BBQ approach models the Josephson junction as a small nonlinear perturbation within a network of linear inductors and capacitors, treating the electromagnetic environment as a ``black box'' characterized by its impedance \cite{PhysRevB.90.134504}. The zeros and poles of the impedance are analyzed to derive Hamiltonian parameters, such as mode frequencies and couplings. This method can also handle multimode systems involving multiple nonlinear modes, making it suitable for complex circuits where the environment significantly affects qubit dynamics. While BBQ provides a comprehensive tool for extracting key Hamiltonian parameters, it requires both the numerical impedance and the junction field distribution. These typically cannot be extracted from the same FEM solver, meaning BBQ requires at least two coordinated simulation steps. This makes it it computationally demanding and dependent on accurate impedance calculations.

\subsubsection{Example of simulation analysis}
The choice of simulation analysis method depends on the specific requirements of the design and the level of precision needed. Each method has its advantages and limitations, and sometimes a combination of methods is employed to leverage their respective strengths. To demonstrate how to perform quantum analysis of FEM simulations, we take the example of the transmon-cavity system described earlier (\cref{sec:SCQ}). We use Ansys as the FEM solver and employ the sub-section simulation methodology. This approach allows us to simulate the transmon and resonator separately, greatly speeding the simulations and making it easy to customize either structure separately. We include the resonator coupling capacitor in both simulations, overlapping the sections to ensure that no inter-section couplings are missed. For the transmon section, we run an electrostatic simulation using Ansys Q3D to obtain the transmon's capacitance matrix, including the coupling capacitor $C_C$ connecting the qubit and the resonator. Using the capacitance matrix and a design value for the junction $E_J$, we can then calculate the transmon frequency and anharmonicity. Similarly, we simulate the resonator subsystem using the Ansys HFSS solver to run an eigenmode simulation. This gives both the resonant frequency $\omega_r$ and linewidth $\kappa$, with the latter determined by coupling to a feedline that we terminate with lumped-element $50 \ \Omega$ resistors in the simulation. The effective resonator capacitance $C_r$ is calculated using $C_r = \frac{\pi}{m \omega_r Z_c}$, where $Z_c$ is the characteristic impedance of the waveguide and $m$ is 2 or 4 for half-wave or quarter-wave designs, respectively. The effective inductance is then $L_r = (C_r \omega_r^2)^{-1}$. Using this lumped-element model, we can then calculate transmon-resonator coupling $g$ either numerically or approximately using the \cref{eq:g}. Similarly, we compute the self-Kerr and cross-Kerr terms either numerically or using \cref{eq:kerr} by combining the results from the EM simulations and the quantum parameters computed from them. Note that we could have used EPR analysis to analyze the whole system and immediately extract all energies (including Kerr terms). This would have required an Ansys HFSS eigenmode simulation of the entire transmon-resonator system, which in our experience can be done accurately but takes much longer.

\section{From Design to Fabrication} \label{sec:fab}

\begin{figure}
    \centering
    \includegraphics[width=\linewidth]{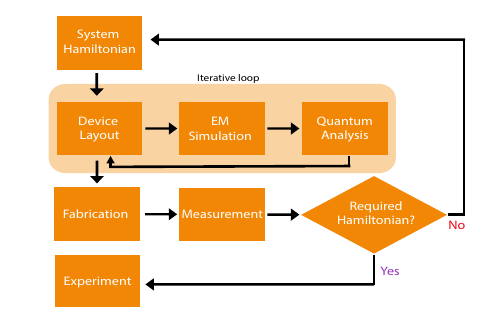}
    \caption{Flow chart of the typical design process. The designer figures out a target system Hamiltonian and circuit parameters. They then make a device layout, feed it to an electromagnetic solver, and analyze the solution to determine whether it will produce the desired Hamiltonian. If the Hamiltonian parameters are out of tolerance, they refine the layout and simulate again, repeating this loop until they match target parameters. Finally they send the design to be fabricated and measured.}
    \label{fig:design-workflow}
\end{figure}

The initial device layout, which is run through EM simulations and subsequent quantum analysis to extract the Hamiltonian parameters, can be created using a variety of tools, including KLayout, Qiskit Metal, KQCircuits, gdsfactory, or even custom design tools developed by individual labs~\cite{KLayout, Qiskit_Metal, kqcircuits, gdsfactory}. While each of these design tools has distinct advantages, we will discuss a design flow using KLayout and Qiskit Metal, as this approach is most commonly used by the community.

Qiskit Metal provides a high-level, Python-based framework that automates the design of superconducting circuits and integrates seamlessly with simulation and quantum analysis workflows. Built specifically with quantum hardware in mind, Qiskit Metal includes pre-defined components, such as qubits, resonators, and couplers, which can be easily configured and connected. Its strong user base, extensive tutorials, and open-source collection of design templates make it a highly accessible and powerful tool. Additionally, Qiskit Metal provides a straightforward API for handling EM simulations and quantum analysis, making it an excellent choice for end-to-end superconducting circuit design. Once a design is created, simulated, and analyzed programmatically within Qiskit Metal, it can be exported to a DXF or GDSII file for further processing - for which we use KLayout; KLayout is widely adopted for its robust and easy-to-use interface for detailed mask design and layout editing. It is particularly popular for its compatibility with the GDSII format, which is essential for mask generation in cleanroom fabrication processes. KLayout also has an extensive Python-based scripting library, enabling custom workflows tailored to the specific requirements of each lab.

Beyond the quantum analysis, the design layout must account for fabrication-specific considerations. For example, if etching is used in the fabrication process, the design needs to include an etch bias (making features wider or narrower as appropriate) to compensate for material loss during the etching step. Additionally, ensuring that components in the design maintain a 50-ohm impedance is crucial, as this minimizes signal reflections and maximizes power transfer---if other impedance is desired, for instance to increase amplifier bandwidth \cite{royBroadbandParametricAmplification2015,qingBroadbandCPWbasedImpedancetransformed2023} or ensure phase-matching conditions in a traveling-wave amplifier \cite{hoeomWidebandLownoiseSuperconducting2012}, this too must be taken into account. Designers must also fine tune the layout to be optimized for their packaging. This could involve decisions such as configuring the design for reflection or transmission measurements based on the number of ports in the package, or strategically placing wirebond pads close to the CPW traces of the printed circuit board (PCB) to ensure proper connections while maintaining sufficient distance from qubits to prevent loss. Once a satisfactory device layout is achieved, multiple copies of the same design are typically fabricated on the sample substrate, limited by the available space, to ensure high yield during fabrication. To further improve the likelihood of a working device, it is recommended to account for fabrication uncertainties within the design itself. For instance, if the desired relationship between the readout resonator and qubit is $\omega_{\text{res}} = 2\omega_q$, and the fabrication process introduces uncertainties in the Josephson inductance $L_j$ within $X\%$, causing qubit frequencies to shift by $\pm Y$ MHz, the design can incorporate several variations of the resonator geometry. This ensures that the resonator frequency $\omega_{\text{res}}$ falls within the range $[2(\omega_q - Y), 2(\omega_q + Y)]$, thereby compensating for potential fabrication-induced variations.

The updated design is re-simulated to verify improvements and ensure that no new issues have been introduced. This iterative process (shown in Fig \ref{fig:design-workflow}) continues until the design satisfies all specifications and is considered ``fab-ready.'' Before finalizing for fabrication, the design undergoes a Design Rule Check (DRC) to ensure it complies with fabrication constraints such as minimum feature sizes, spacing, layer overlap, etc. Any violations flagged during the DRC stage must be reviewed and either corrected or explicitly waived with justification, to avoid unintended fabrication failures. After fabrication, the device is packaged, cooled, and measured as described in \cref{sec:validation}. Hamiltonian parameters are extracted to validate the simulation pipeline. Discrepancies between the measured and simulated results can arise due to inaccuracies in material models, parasitic effects, or variations during the fabrication process. These discrepancies are analyzed by comparing the required and observed Hamiltonian parameters, leading to refinements in the simulation models by adjusting material properties, re-calibrating simulation hyperparameters, or revising geometric tolerances. This feedback loop is crucial to developing robust simulation workflows and ensuring that fabricated devices perform as expected.

High-accuracy simulations are often resource- and time-intensive, making them less accessible for smaller research groups and creating bottlenecks in the design-to-fabrication process. Seamless integration between EM simulations and quantum parameter extraction methods is essential for efficient workflows but is not always easily achievable with existing tools. Initiatives like SQuADDS (\textbf{S}uperconducting \textbf{Qu}bit \textbf{A}nd \textbf{D}evice \textbf{D}esign and \textbf{S}imulation database)~\cite{Shanto2024squaddsvalidated} address these challenges by providing a platform for sharing validated simulation pipelines, measured device results, and designs. SQuADDS offers tools that automate the design, simulation, and analysis process, enabling the rapid generation of validated, fabrication-ready device layouts. By providing access to a database of proven designs, SQuADDS reduces the need for redundant simulations and accelerates the design-to-fabrication workflow across the community.

\section{Closing the design loop with measurements}\label{sec:validation}
After a design has been fabricated, it is important to check whether it accomplished its goals: are the mode frequencies, anharmonicities, and couplings correct? Are readout resonator linewidths appropriate? Are quantum and classical crosstalk sufficiently suppressed? Are coherences sufficiently high? Each of these points must be addressed in order. Mode frequencies and anharmonicities are often measured spectroscopically, as are resonator linewidths. Coherence times may be measured using standard population decay, Ramsey, and Hahn echo experiments. Mode couplings can be measured several ways: tuning one mode through another and measuring the splitting of the hybridized states that result when the two modes are on resonance; tuning modes into resonance and measuring the rate of energy exchange in the time domain; changing the state of one mode and looking for a cross-Kerr shift in the frequency of another mode (in this case the mode frequencies are also needed to calculate coupling strength); and ``punching out'' a qubit by driving it so hard that it escapes the Josephson potential well, effectively removing it from the circuit, and then measuring the Lamb shift on other modes \cite{finkQuantumClassicalTransitionCavity2010}. Quantum crosstalk can be measured with similar spectroscopic and cross-Kerr measurement techniques, while classical crosstalk is best characterized by measuring how easily a mode can be driven using lines that are supposedly not coupled to that mode \cite{yanCalibrationCancellationMicrowave2023}.

One device properties have been characterized, it is important to determine whether any discrepancies with the target behavior are due to device design, fabrication, or embedding. Coherence measurements of control devices with standard designs can determine whether design is to blame for reduced coherence or whether it can be attributed to fabrication issues and/or improper filtering of measurement lines and/or improper shielding. Optical measurements can determine if the device geometry was fabricated as expected. Room-temperature resistance measurements of test structures can determine if Josephson energies were as designed. Note that junction behavior can include departure from the single cosinusoidal energy expected for an ideal tunnel junction, which may alter spectroscopic results but not room-temperature resistance. Such behavior must be observed with low-temperature measurements \cite{willschObservationJosephsonHarmonics2024}. 

Once fabrication issues are eliminated, incorrect mode frequencies and couplings, excess crosstalk, and unwanted resonances can be confidently attributed to design. The most common reason for these missed parameters is an inaccurate electromagnetic simulation of the device and/or an inaccurate translation of simulation results to Hamiltonian. Inaccurate simulation can be caused by, e.g., a simulation that does not include the whole geometry and thus fails to capture some long-range coupling or distributed mode; a simulation that does not have a tight enough tolerance for convergence or a fine enough mesh and thus produces an inaccurate solution; a simulation that fails to take into account the real geometry implemented including fabrication biases; or a simulation that does not take into account significant kinetic inductance. Inaccurate interpretation can be caused by, e.g. an interpretation that fails to account for finite-size effects of ``lumped'' elements that are not small compared to the wavelength; an interpretation that improperly treats distributed elements; an interpretation that truncates the basis space at too small of a dimension; an interpretation that uses approximate analytical formulas beyond their regime of accuracy; or an interpretation that fails to account for quantum effects such as phase slips. Using a robust and validated simulation pipeline is thus crucial for any device designer. Again the usual approach is empirical: if experimental results disagree with simulation, designers run larger simulations with tighter tolerances and add complexity to their interpretation models until agreement is reached.

\section{Outlook} \label{sec:conclusion}
As the complexity of superconducting quantum circuit devices grows, so too does the difficulty of designing them. At the same time, the resources required to fabricate and measure a device are also growing, making it even more crucial to correctly design the device the first time. Likewise, centralized foundries are beginning to replace in-house fabrication for many groups, so that even simple devices may have long fabrication turnaround times. There is need for further development of tools to aid designers. Fortunately, open-source community resources are rapidly growing. Qiskit Metal and KQCircuits provide automated circuit layout. Scqubits and SQCircuit allow straightforward calculation of circuit properties. SQuADDS provides ready-made designs for standard devices. Further development of these tools can aid the community, as can development of open-source automated design rule check (DRC) software, faster and more computationally-efficient simulation techniques and software, and standardized design best practices.

Similarly, increasing device complexity means it is more difficult to \emph{learn} to design devices. A layout designer must have some understanding of the physics that their target circuit implements, otherwise they would not know which stray couplings may be acceptable and which would hinder performance. Likewise, a circuit graph designer must know what kind of layouts are feasible, and how different components can affect device coherence and crosstalk. To date, most of this knowledge is passed down through research groups, with occasional publications and talks explaining how one aspect of the problem works. There is a need for pedagogical reference materials, including textbooks, practical tutorials, and interactive demonstrations, and for dedicated courses on device design. Many such materials and courses have already been developed for quantum computing from an algorithmic perspective (``top down''), and the time has come to extend to hardware design (``bottom up'').

Each topic covered in this short review is worthy of far more discussion, and the papers referenced here cover them in great detail. In \cref{tab:references} we collect some of the key references, organized by topic. It is our hope that, until extensive didactic materials are developed, this review will function as a quick introduction for new designers and a central index for experienced researchers looking to dig more into one topic. 

\begin{acknowledgements}
We gratefully acknowledge useful discussions with Jens Koch, Vivek Maurya, Zlatko Minev, and Hakan T{\"u}reci. Some devices discussed were fabricated and provided by the Superconducting Qubits at Lincoln Laboratory (SQUILL) Foundry at MIT Lincoln Laboratory, with funding from the Laboratory for Physical Sciences (LPS) Qubit Collaboratory. Funding was provided by the National Science Foundation Quantum Leap Big Idea under Grant No. OMA-1936388, the Office of Naval Research under Grant No. N00014-21-1-2688, the Air Force Office of Scientific Research under Grant No. FA9550-23-1-0165, the Defense Advanced Research Projects Agency under MeasQUIT HR0011-24-9-0362, and Research Corporation for Science Advancement under Cottrell Scholar Grant 27550.
\end{acknowledgements}

\pagebreak

\onecolumngrid
\begin{longtable}{|l|l|l|l|}
\caption{References ordered by topic, with reviews and resources at the top and then topics listed in order of appearance in the text. Some references unrelated to designs are omitted. A * next to an author name indicates a review article.}
\label{tab:references}\\
\hline
Topic                                                                     & \multicolumn{1}{c|}{Lead Author} & \multicolumn{1}{c|}{Year} & Reference \\ \hline
\endfirsthead
\endhead
\multirow{9}{*}{General superconducting qubits reviews}                   & Clarke*                          & 2008                      &    \cite{clarkeSuperconductingQuantumBits2008}       \\ \cline{2-4} 
                                               & Devoret*    & 2013 &  \cite{devoretSuperconductingCircuitsQuantum2013a}\\ \cline{2-4} 
                                               & Wendin*     & 2017 &  \cite{wendinQuantumInformationProcessing2017a}\\ \cline{2-4} 
                                               & Krantz*     & 2019 & \cite{krantzQuantumEngineersGuide2019} \\ \cline{2-4} 
                                               & Kjaergaard* & 2020 & \cite{kjaergaardSuperconductingQubitsCurrent2020} \\ \cline{2-4} 
                                               & Kwon*       & 2021 & \cite{kwonGatebasedSuperconductingQuantum2021} \\ \cline{2-4} 
                                               & Blais*      & 2021 & \cite{blaisCircuitQuantumElectrodynamics2021a} \\ \cline{2-4} 
                                               & Gao*        & 2021 & \cite{gaoPracticalGuideBuilding2021} \\ \cline{2-4} 
                                               & Rasmussen*  & 2021 & \cite{rasmussenSuperconductingCircuitCompanionan2021} \\ \hline
\multirow{3}{*}{Cavity qubits reviews}         & Joshi*      & 2021 & \cite{joshiQuantumInformationProcessing2021} \\ \cline{2-4} 
                                               & Ma*         & 2021 & \cite{maQuantumControlBosonic2021} \\ \cline{2-4} 
                                               & Krasnok*    & 2024 & \cite{krasnokSuperconductingMicrowaveCavities2024} \\ \hline
\multirow{10}{*}{Open-source codes}            & Malinen     & 2013 & \cite{Malinen2013} \\ \cline{2-4} 
                                               & Grozkowski  & 2021 & \cite{groszkowskiScqubitsPythonPackage2021a} \\ \cline{2-4} 
                                               & Minev       & 2021 & \cite{Qiskit_Metal} \\ \cline{2-4} 
                                               & Minev       & 2021 & \cite{pyEPR} \\ \cline{2-4} 
                                               & Cucurachi   & 2021 & \cite{kqcircuits} \\ \cline{2-4} 
                                               & Aumann      & 2022 & \cite{aumannCircuitQOpensourceToolbox2022} \\ \cline{2-4} 
                                               & Rajabzadeh  & 2023 & \cite{rajabzadehAnalysisArbitrarySuperconducting2023} \\ \cline{2-4} 
                                               & Choi        & 2023 & \cite{Choi2023} \\ \cline{2-4} 
                                               & Shanto      & 2024 & \cite{Shanto2024squaddsvalidated} \\ \cline{2-4} 
                                               & Shammah*    & 2024 & \cite{Shammah_2024} \\ \hline
\multirow{13}{*}{Circuit quantization and diagonalization}                 & Ulrich                           & 2016                      &    \cite{ulrichDualApproachCircuit2016}       \\ \cline{2-4} 
                                               & Vool*       & 2017 & \cite{voolIntroductionQuantumElectromagnetic2017} \\ \cline{2-4} 
                                               & Solgun         & 2019 & \cite{solgunSimpleImpedanceResponse2019a} \\ \cline{2-4} 
                                               & Kerman      & 2020 &  \cite{kermanEfficientNumericalSimulation2020} \\ \cline{2-4} 
                                               & Chitta      & 2022 & \cite{chittaComputeraidedQuantizationNumerical2022} \\ \cline{2-4} 
                                               & Solgun      & 2022 & \cite{solgunDirectCalculation$ZZ$2022} \\ \cline{2-4} 
                                               & Egusquiza      & 2022 & \cite{egusquizaAlgebraicCanonicalQuantization2022} \\ \cline{2-4} 
                                               & Rajabzadeh  & 2023 & \cite{rajabzadehAnalysisArbitrarySuperconducting2023} \\ \cline{2-4}
                                               & Osborne  & 2024 & \cite{osborneSymplecticGeometryCircuit2024} \\ \cline{2-4}
                                               & Parra-Rodriguez  & 2024 & \cite{parra-rodriguezGeometricalDescriptionFaddeevJackiw2024} \\ \cline{2-4}
                                               & Weissler    & 2024 & \cite{weisslerEnumerationAllSuperconducting2024} \\ \cline{2-4}
                                               & Labarca    & 2024 & \cite{labarcaToolboxNonreciprocalDispersive2024} \\ \cline{2-4}
                                               & Parra-Rodriguez    & 2024 & \cite{parra-rodriguezExactQuantizationNonreciprocal2025} 
                                               \\ \hline
\multirow{3}{*}{Quantizing in the presence of time-dependent flux}                                                & You    & 2019 & \cite{youCircuitQuantizationPresence2019} \\ \cline{2-4}
                                               & Riwar    & 2022 & \cite{riwarCircuitQuantizationTimedependent2022} \\ \cline{2-4}
                                               & Bryon    & 2023 & \cite{bryonTimeDependentMagneticFlux2023} 
                                               \\ \hline
\multirow{5}{*}{Constructing circuit graphs and effective Hamiltonians from simulation}   & Solgun                           & 2014                      &   \cite{PhysRevB.90.134504}        \\ \cline{2-4} 
                                               & Besedin     & 2018 & \cite{besedinQualityFactorTransmission2018} \\ \cline{2-4} 
                                               & Minev       & 2021 & \cite{Minev2021} \\ \cline{2-4} 
                                               & Minev       & 2021 & \cite{minev2021circuitquantumelectrodynamicscqed} \\ \cline{2-4} 
                                               & Yuan        & 2022 & \cite{Yuan2022} \\ \hline
\multirow{6}{*}{Effects not captured by the rotating wave approximation}               & Zueco                            & 2009                      &     \cite{zuecoQubitoscillatorDynamicsDispersive2009}      \\ \cline{2-4} 
                                               & Sank        & 2016 & \cite{sankMeasurementInducedStateTransitions2016a} \\ \cline{2-4} 
                                               & Dumas       & 2024 & \cite{dumasMeasurementInducedTransmonIonization2024} \\ \cline{2-4} 
                                               & Scheuer     & 2014 & \cite{scheuerPreciseQubitControl2014} \\ \cline{2-4} 
                                               & Willsch     & 2017 & \cite{willschGateerrorAnalysisSimulations2017} \\ \cline{2-4} 
                                               & Ann         & 2021 & \cite{annSidebandTransitionsTwomode2021} \\ \hline
\multirow{7}{*}{Classical crosstalk}           & Abrams      & 2019 & \cite{abramsMethodsMeasuringMagnetic2019} \\ \cline{2-4} 
                                               & Dai         & 2021 & \cite{daiCalibrationFluxCrosstalk2021} \\ \cline{2-4} 
                                               & Winick      & 2021 & \cite{winickSimulatingMitigatingCrosstalk2021} \\ \cline{2-4} 
                                               & Wang        & 2022 & \cite{wangControlMitigationMicrowave2022} \\ \cline{2-4} 
                                               & Ketterer    & 2023 & \cite{kettererCharacterizingCrosstalkSuperconducting2023} \\ \cline{2-4} 
                                               & Yan         & 2023 & \cite{yanCalibrationCancellationMicrowave2023} \\ \cline{2-4} 
                                               & Yang        & 2024 & \cite{yangFastUniversalScheme2024} \\ \hline
\multirow{7}{*}{Quantum crosstalk}             & Sheldon     & 2016 & \cite{sheldonProcedureSystematicallyTuning2016} \\ \cline{2-4} 
                                               & Yan         & 2018 & \cite{yanTunableCouplingScheme2018} \\ \cline{2-4} 
                                               & Mundala     & 2019 & \cite{mundadaSuppressionQubitCrosstalk2019} \\ \cline{2-4} 
                                               & Mitchell    & 2021 & \cite{mitchellHardwareEfficientMicrowaveActivatedTunable2021} \\ \cline{2-4} 
                                               & Kandala     & 2021 & \cite{kandalaDemonstrationHighFidelityCnot2021} \\ \cline{2-4} 
                                               & Wei         & 2022 & \cite{weiHamiltonianEngineeringMulticolor2022} \\ \cline{2-4} 
                                               & Tripathi    & 2022 & \cite{tripathiSuppressionCrosstalkSuperconducting2022} \\ \hline
\multirow{7}{*}{Air-bridge crossovers}         & Abuwasib    & 2013 & \cite{abuwasibFabricationLargeDimension2013} \\ \cline{2-4} 
                                               & Chen        & 2014 & \cite{chenFabricationCharacterizationAluminum2014} \\ \cline{2-4} 
                                               & Dunsworth   & 2018 & \cite{dunsworthMethodBuildingLow2018} \\ \cline{2-4} 
                                               & Sun         & 2022 & \cite{sunFabricationAirbridgesGradient2022} \\ \cline{2-4} 
                                               & Janzen      & 2022 & \cite{janzenAluminumAirBridges2022} \\ \cline{2-4} 
                                               & Tao         & 2024 & \cite{taoFabricationCharacterizationLow2024} \\ \cline{2-4} 
                                               & Bu          & 2024 & \cite{buTantalumAirbridgesScalable2024} \\ \hline
\multirow{7}{*}{Flip-chip and 3D integration}  & Rosenberg   & 2017 & \cite{rosenberg3DIntegratedSuperconducting2017}  \\ \cline{2-4} 
                                               & Foxen       & 2017 & \cite{foxenQubitCompatibleSuperconducting2017} \\ \cline{2-4} 
                                               & Vahidpour   & 2017 & \cite{vahidpourSuperconductingSiliconVias2017} \\ \cline{2-4} 
                                               & Yost        & 2020 & \cite{yostSolidstateQubitsIntegrated2020a} \\ \cline{2-4} 
                                               & Rosenberg*  & 2020 & \cite{rosenbergSolidStateQubits3D2020a} \\ \cline{2-4} 
                                               & Li       & 2021 & \cite{liVacuumgapTransmonQubits2021} \\ \cline{2-4} 
                                               & Kosen       & 2024 & \cite{kosenSignalCrosstalkFlipChip2024} \\ \hline
Connectivity and error correction              & Gottesman   & 2014 & \cite{gottesmanFaultTolerantQuantumComputation2014} \\ \hline
\multirow{4}{*}{Short-range coupling schemes}                             & Moskalenko                       & 2022                      &     \cite{moskalenkoHighFidelityTwoqubit2022}      \\ \cline{2-4} 
                                               & Weiss       & 2022 & \cite{weissFastHighFidelityGates2022} \\ \cline{2-4} 
                                               & Ding        & 2023 & \cite{dingHighFidelityFrequencyFlexibleTwoQubit2023} \\ \cline{2-4} 
                                               & Zhang       & 2024 &  \cite{zhangTunableInductiveCoupler2024} \\ \hline
\multirow{8}{*}{Long-range coupling schemes}   & Roch        & 2014 & \cite{rochObservationMeasurementinducedEntanglement2014} \\ \cline{2-4} 
                                               & Schwartz    & 2016 & \cite{kimchi-schwartzStabilizingEntanglementSymmetrySelective2016} \\ \cline{2-4} 
                                               & Kurpiers    & 2018 & \cite{kurpiersDeterministicQuantumState2018} \\ \cline{2-4} 
                                               & Leung       & 2019 & \cite{leungDeterministicBidirectionalCommunication2019} \\ \cline{2-4} 
                                               & Kannan      & 2023 & \cite{kannanDemandDirectionalMicrowave2023} \\ \cline{2-4} 
                                               & Zhou        & 2023 & \cite{zhouRealizingAllallCouplings2023} \\ \cline{2-4} 
                                               & Greenfield  & 2024 & \cite{greenfieldStabilizingTwoqubitEntanglement2024} \\ \cline{2-4}
                                               & Almanakly   & 2025 & \cite{almanaklyDeterministicRemoteEntanglement2025} \\ \hline
\multirow{7}{*}{Dielectric loss}               & Gao         & 2008 & \cite{gaoExperimentalEvidenceSurface2008} \\ \cline{2-4} 
                                               & Paik        & 2011 & \cite{paikObservationHighCoherence2011} \\ \cline{2-4} 
                                               & Oliver*     & 2013 & \cite{oliverMaterialsSuperconductingQuantum2013a} \\ \cline{2-4} 
                                               & Wang        & 2015 & \cite{wangSurfaceParticipationDielectric2015} \\ \cline{2-4} 
                                               & Siddiqi*    & 2021 & \cite{siddiqiEngineeringHighcoherenceSuperconducting2021b} \\ \cline{2-4} 
                                               & Martinis    & 2022 & \cite{martinisSurfaceLossCalculations2022a} \\ \cline{2-4} 
                                               & Smirnov     & 2024 & \cite{smirnovWiringSurfaceLoss2024} \\ \hline
\multirow{8}{*}{Radiation and quasiparticles}  & Barends     & 2011 & \cite{barendsMinimizingQuasiparticleGeneration2011a} \\ \cline{2-4} 
                                               & Hays        & 2018 & \cite{haysDirectMicrowaveMeasurement2018} \\ \cline{2-4} 
                                               & Houzet      & 2019 & \cite{houzetPhotonAssistedChargeParityJumps2019} \\ \cline{2-4} 
                                               & Rafferty    & 2021 & \cite{raffertySpuriousAntennaModes2021a} \\ \cline{2-4} 
                                               & Farmer      & 2021 & \cite{farmerContinuousRealtimeDetection2021} \\ \cline{2-4} 
                                               & Iaia        & 2022 & \cite{iaiaPhononDownconversionSuppress2022} \\ \cline{2-4} 
                                               & Elfeky      & 2023 & \cite{elfekyQuasiparticleDynamicsEpitaxial2023a} \\ \cline{2-4} 
                                               & Liu         & 2024 & \cite{liuQuasiparticlePoisoningSuperconducting2024} \\ \hline
Packaging                                      & Huang       & 2021 & \cite{huangMicrowavePackageDesign2021} \\ \hline
\multirow{4}{*}{Signal injection in 3D cavities} & Axline      & 2016 & \cite{axlineArchitectureIntegratingPlanar2016} \\ \cline{2-4} 
                                               & Reshitnyk   & 2016 & \cite{reshitnyk3DMicrowaveCavity2016} \\ \cline{2-4} 
                                               & Stammeier   & 2018 & \cite{stammeierApplyingElectricMagnetic2018} \\ \cline{2-4} 
                                               & Gargiulo    & 2021 & \cite{gargiuloFastFluxControl2021} \\ \hline
\multirow{6}{*}{Electromagnetic simulation}    & Zmuidzinas* & 2012 & \cite{Zmuidzinas2012} \\ \cline{2-4} 
                                               & Kurra       & 2014 & \cite{fem_mesh_paper} \\ \cline{2-4} 
                                               & Naaman      & 2024 & \cite{naaman2024modelingfluxquantizingjosephsonjunction} \\ \cline{2-4} 
                                               & Hauer       & 2024 & \cite{Hauer2024} \\ \cline{2-4} 
                                               & Pham        & 2023 & \cite{phamFluxbasedThreedimensionalElectrodynamic2023} \\ \cline{2-4} 
                                               & Pham        & 2024 & \cite{phamSpectralTheoryNonlinear2024} \\ \hline
\multirow{2}{*}{Characterization measurements} & Fink        & 2010 & \cite{finkQuantumClassicalTransitionCavity2010} \\ \cline{2-4} 
                                               & Willsch     & 2024 & \cite{willschObservationJosephsonHarmonics2024} \\ \hline
\end{longtable}

\twocolumngrid

\bibliography{references}
\end{document}